\documentclass[aps,
twocolumn,
pra,
superscriptaddress]{revtex4-1}
\usepackage{graphicx}
\usepackage{color}
\usepackage{ulem}
\usepackage{verbatim}
\usepackage{amsmath}
\usepackage{amsfonts}
\usepackage{amssymb}

\usepackage{hyperref}
\hypersetup{
  colorlinks   = true, 
  urlcolor     = blue, 
  linkcolor    = blue, 
  citecolor   = blue 
}

\usepackage{array}
\usepackage{tabulary}
\newcolumntype{K}[1]{>{\centering\arraybackslash}p{#1}}

\newcommand{\ba}{\begin{eqnarray}}
\newcommand{\ea}{\end{eqnarray}}
\newcommand{\ban}{\begin{eqnarray*}}
\newcommand{\ean}{\end{eqnarray*}}

\newcommand{\ket}[1]{\mbox{$ | #1 \rangle $}}

\begin{document}

\title{Non-classical correlations between a C-band telecom photon and a stored spin-wave}%
\pacs{42.50.Dv, 03.67.Hk, 32.80.Qk}



\author{Pau Farrera}
\altaffiliation{These authors contributed equally to this work.}
\affiliation{ICFO-Institut de Ciencies Fotoniques, The Barcelona Institute of Science and Technology, 08860 Castelldefels (Barcelona), Spain}

\author{Nicolas Maring}
\altaffiliation{These authors contributed equally to this work.}
\affiliation{ICFO-Institut de Ciencies Fotoniques, The Barcelona Institute of Science and Technology, 08860 Castelldefels (Barcelona), Spain}

\author{Boris Albrecht}
\altaffiliation{Present address: Niels Bohr Institute, University of Copenhagen, Denmark}
\affiliation{ICFO-Institut de Ciencies Fotoniques, The Barcelona Institute of Science and Technology, 08860 Castelldefels (Barcelona), Spain}

\author{Georg Heinze}
\email[Contact: ]{georg.heinze@icfo.es}
\affiliation{ICFO-Institut de Ciencies Fotoniques, The Barcelona Institute of Science and Technology, 08860 Castelldefels (Barcelona), Spain}

\author{Hugues de Riedmatten}
\homepage{http://qpsa.icfo.es}
\affiliation{ICFO-Institut de Ciencies Fotoniques, The Barcelona Institute of Science and Technology, 08860 Castelldefels (Barcelona), Spain}
\affiliation{ICREA-Instituci\'{o} Catalana de Recerca i Estudis Avan\c cats, 08015 Barcelona, Spain}%


\begin{abstract}
Future ground-based quantum information networks will likely use single photons transmitted through optical fibers to entangle individual network nodes. To extend communication distances and overcome limitations due to photon absorption in fibers the concept of quantum repeaters has been proposed. For that purpose, it is required to achieve quantum correlations between the material nodes and photons at telecom wavelengths which can be sent over long distances in optical fibers. Here we demonstrate non-classical correlation between a frequency converted telecom C-band photon and a spin-wave stored in an atomic ensemble quantum memory. The photons emitted from the ensemble and heralding the spin-waves  are converted from 780~nm to 1552~nm by means of an all-solid-state integrated waveguide non-linear device. We show ultra-low noise operation of the device enabling  a high signal to noise ratio of the converted single photon, leading to a high spin-wave heralding efficiency. The presented work is an enabling step towards the practical entanglement of remote quantum memories and the entanglement of quantum systems operating at different wavelengths.
\end{abstract}

\maketitle

\section{Introduction}
Long distance quantum communication has been an ambitious and long standing goal in the quantum information community. Although commercial solutions are available for short distance quantum key distribution, the extension to large scales remains challenging due to the exponential scaling of photon losses in optical fibers with distance \cite{Korzh2015}. To overcome this limitation the concept of quantum repeaters (QR) has been developed which holds great promise to extend the operation distance towards continental scale \cite{Briegel1998,Sangouard2011}. 

The essential building block of most quantum repeater achitectures  is a photonic quantum memory (QM) which provides an interface between stationary quantum bits (encoded in atom-like systems) and flying quantum bits (encoded in photons) \cite{Afzelius2015}. QMs have been implemented in many different systems such as single atoms and ions, atomic ensembles, and various solid state systems \cite{Lvovsky2009,Hammerer2010,Simon2010,Sangouard2011,Bussieres2013,Afzelius2015}. Heralded entanglement between remote QMs can be achieved by quantum interference of photonic modes correlated to the QMs at a central measurement station \cite{Chou2005,Chou2007,Yuan2008,Moehring2007,Hensen2015}. To achieve long distance entanglement, it is essential that the heralding photon be at telecom wavelength, in order to minimize loss in optical fibers, which has not been the case so far in previous experiments. A crucial enabling step is therefore the ability to obtain quantum correlation between a telecom photon (preferably in the C band where the loss is minimal) and a long-lived atomic state in a quantum memory.  

The best current QMs operate with photons in the visible or near infrared regime \cite{Lvovsky2009,Hammerer2010,Simon2010,Sangouard2011,Bussieres2013,Afzelius2015}, which strongly limits the possibilities of long distance transmission. Although progress is being made towards quantum memories functioning directly at telecom wavelengths \cite{Lauritzen2010,Dajczgewand2014,Saglamyurek2015}, the current demonstrations in the quantum regime still suffer from short coherence times (ns time-scale) and low efficiencies. There are basically two different approaches to connect visible QMs to telecom wavelengths. The first one is to use photon pair sources intrinsically emitting entangled photon pairs with one photon at telecom wavelength and the other one in the visible or near infrared regime to be memory compatible. This photon then can be stored in a read-write QM where the photon is mapped onto an atomic state and can be retrieved back on demand  \cite{Simon2007,Clausen2011,Saglamyurek2011,Rielander2014,Bussieres2014,Lenhard2015,Schunk2015,Zhang2016}. The second possibility is to use read QMs emiting photons entangled with the atoms, and to convert the emitted photons to telecom wavelengths. Significant efforts have been devoted to quantum frequency conversion (QFC) of single photons towards the telecom band \cite{Tanzilli2005,Radnaev2010,Dudin2010,Ikuta2011,Zaske2012,Ates2012,Ikuta2013,Albrecht2014,Maring2014}, although only very few examples are compatible with quantum memories \cite{Radnaev2010,Dudin2010,Albrecht2014,Maring2014}. Ref.~\cite{Albrecht2014} demonstrated the quantum frequency conversion to the C-band of a single read-out photon emitted by a Rb-based quantum memory with an efficiency of $30\,\%$, using an integrated waveguide approach. However, in that experiment there was no correlation left between the converted photon and the quantum memory. The conversion of the heralding photon from a QM was so far only realized in a single experiment in which photons were converted form $795\,\mathrm{nm}$ to $1367\,\mathrm{nm}$ (E-band) via four wave mixing (FWM) in a cold and dense ensemble of Rubidium atoms \cite{Radnaev2010,Dudin2010}. In contrast to the conversion of the read-out photon, the heralding write photon conversion is much more challenging in terms of noise suppression, because the emission probability of the write photon needs to be low (typ. $<1\,\%$) to avoid multiple spin-wave-excitations in the same mode leading to uncorrelated photons emitted in that mode. Hence, a constant background-noise due to the conversion process has a much higher impact on the SNR of the heralding write photons than of the heralded read photons. In order to reach a high heralding efficiency after conversion, the background noise must therefore be very low.

In the present paper, we demonstrate low-noise quantum frequency conversion of the initial heralding write photon from a cold Rubidium QM to $1552\,\mathrm{nm}$ via difference frequency generation in an all-solid-state integrated non-linear waveguide. The QM is implemented in a cold ensemble of $^{87}\mathrm{Rb}$ atoms following the scheme of Duan, Lukin, Cirac, and Zoller (DLCZ) \cite{Duan2001} and we use a periodically poled lithium niobate (PPLN) waveguide for conversion. In contrast to the FWM approach in cold atoms, QFC via solid-state waveguides offers major advantages such as wavelength flexibility, robustness, relative simplicity, and excellent prospects for on-chip-integration. By combining high QFC efficiency and ultra-narrowband filtering, we demonstrate that with the presented approach a high degree of non-classical correlations between an atomic spin-wave stored in the QM and a flying telecom photon can be achieved, as well as high signal-to-noise ratio (SNR) for the detection of the converted heralding photon, leading to a  high spin-wave heralding efficiency.

\section{Implementation and Setup}
\label{sec:Implementation}

\begin{figure}
\centering
\includegraphics[width=\linewidth]{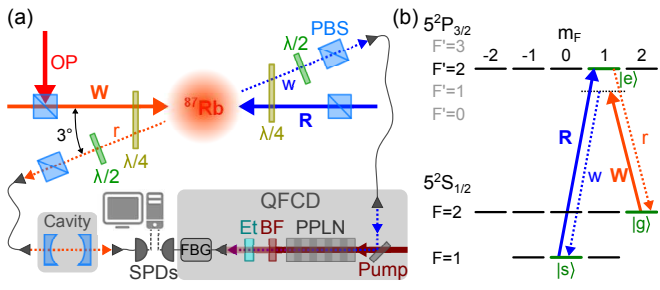}
\caption{(color online) (a) Experimental setup. Write pulse ($W$) and Read pulse ($R$) are sent counter-propagating into the atomic cloud. Write and read photonic modes are denoted by $w$ and $r$. The QFCD consists of the PPLN waveguide, a bandpass filter (BF), a narrowband etalon (Et), and a fiber bragg grating (FBG). (b) Energy levels and coupling scheme for the DLCZ experiment.}
\label{Figure1}
\end{figure}

The experimental setup is depicted in Fig.~\ref{Figure1}(a) and basically consists of two parts -- the atomic QM and the quantum frequency conversion device (QFCD). After cooling the $^{87}\mathrm{Rb}$ atoms in a magneto optical trap (MOT) they are prepared in the ground state $|g\rangle = |5^2S_{1/2},F=2,m_F=2\rangle$ by optical pumping (cf. Fig.~\ref{Figure1}(b)). A weak Gaussian-shaped write pulse (FWHM duration $\tau_W=20\,\mathrm{ns}$), $40\,\mathrm{MHz}$ red-detuned from the $|g\rangle \leftrightarrow |e\rangle = |5^2P_{3/2},F'=2,m_F=1\rangle$ transition, probabilistically creates a single collective spin excitation (spin-wave) between the two ground states $|g\rangle$ and $|s\rangle = |5^2S_{1/2},F=1,m_F=0\rangle$. The spin-wave can be stored for a programmable time in the QM and is heralded by a Raman scattered write photon ($\tau_w=22\,\mathrm{ns}$). The write photon is circularly polarized with respect to the experiments' quantization axis, set by a weak homogeneous static magnetic field applied over the whole cloud. We couple a small fraction of the isotropically emitted write photons under an angle of $3^{\circ}$ with respect to the write pulse axis into a single mode fiber (coupling efficiency approx. $60\%$). Besides that spatial filtering, we also perform polarization filtering of the write photon using a combination of quarter- and half-waveplates, as well as a polarization beam splitter in front of the fiber. 

The afterwards linearly polarized write photon is sent to the QFCD and first overlapped on a dichroic mirror with the spatial mode of the pump laser at $1569\,\mathrm{nm}$ which before was spectrally cleaned by two bandpass filters (Semrock NIR1, center wavelength $1570\,\mathrm{nm}$, transmission bandwidth $8.9\,\mathrm{nm}$), leading to an ASE suppression of more than $100\,\mathrm{dB}$ at $1552\,\mathrm{nm}$. A combination of lenses (not shown in Fig.~\ref{Figure1}(a)) ensures optimal focussing and mode matching of the beams into the temperature stabilized $3\,\mathrm{cm}$ long PPLN waveguide (HC Photonics) in which the conversion of the write photon from $780\,\mathrm{nm}$ to $1552\,\mathrm{nm}$ takes place. Afterwards, the pump radiation is blocked by a combination of two bandpass filters (Semrock NIR01-1550/3-25) each with a transmission bandwidth of $7\,\mathrm{nm}$ around $1552\,\mathrm{nm}$ and an maximum optical depth of $OD \approx 12$. However, further filtering is required to detect the converted write photon at the single photon level because of noise generated by spontaneous Raman scattering of the pump beam which leads to a broad background around the target wavelength. In contrast to former work \cite{Albrecht2014,Fernandez-Gonzalvo2013}, we apply a two-stage additional filtering consisting of an etalon with a bandwidth of $210\,\mathrm{MHz}$ and a free spectral range of $4\,\mathrm{GHz}$ and a fiber Bragg grating (FBG) of $2.5\,\mathrm{GHz}$ bandwidth. The total extinction ratio of the whole filtering stage for the pump radiation at $1569\,\mathrm{nm}$ is $>\!\!150\,\mathrm{dB}$ ($100\,\mathrm{dB}$ for the two bandpass filters, $44\,\mathrm{dB}$ for the FBG and $11\,\mathrm{dB}$ for the etalon). This allowed us to achieve high values of SNRs at low photon number, which is necessary for the quantum frequency conversion of the heralding write single photons. The converted write photons are finally detected by an InGaAs single photon detector (SPD) (ID Quantique ID230) with an detection efficiency of $\eta_{d,1552}=10\%$.

\section{Results}
\label{sec:Results}

\subsection{QFCD performance}
\label{sec:QFCD}

\begin{figure}
\centering
\includegraphics[width=\linewidth]{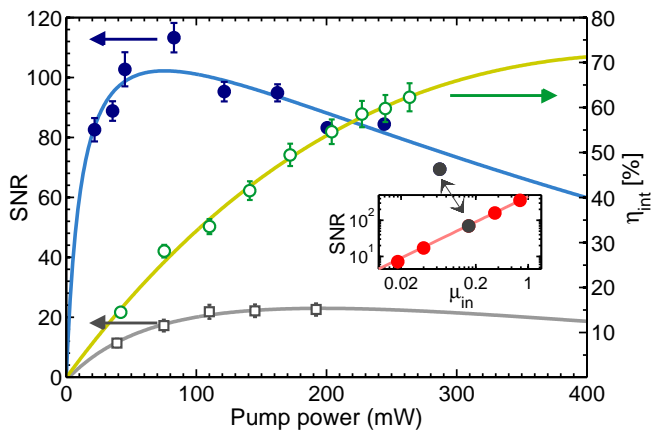}
\caption{(color online) Signal to noise ratio SNR (blue dots for full filtering, grey squares without etalon, left axis) measured with a mean input photon number per pulse of $\mu_\mathrm{in}=0.16$ and internal efficiency $\eta_\mathrm{int}$ of the QFCD (green circles, right axis) measured with classical input light vs. pump power measured after the waveguide. The data are fitted by functions, modeling the expected behavior (solid lines). The inset shows the SNR vs. $\mu_\mathrm{in}$ for a fixed pump power of $P_\mathrm{pump}=287\,\mathrm{mW}$.}
\label{Figure2}
\end{figure}

The performance of the QFCD can be deduced from Figure~\ref{Figure2}. For characterization of the QFCD, we couple $1.2\,\mathrm{mW}$ continuous wave input light at $780\,\mathrm{nm}$ to measure the total conversion efficiency and single photon level coherent input pulses of $16\,\mathrm{ns}$ duration with a mean photon number per pulse of $\mu_\mathrm{in}=0.16$ to measure the SNR versus the coupled pump power (measured behind the waveguide). The plotted internal efficiency $\eta_\mathrm{int}$ excludes all optical losses, e.g. due to initial coupling in the waveguide ($\eta_\mathrm{cpl}\approx 74\%$), all subsequent filtering stages ($\eta_\mathrm{filter}\approx 36\%$), all optical surfaces including one optical isolator ($\eta_\mathrm{surf}\approx 70\%$) and the final fiber coupling ($\eta_\mathrm{fiber}\approx 75\%$). The data are fitted with the models described in \cite{Albrecht2014,Fernandez-Gonzalvo2013} and we retrieve a normalized conversion efficiency of $\eta_{n}=61\,\%/\mathrm{W}/\mathrm{cm}^2$ and a maximum internal efficiency of $\eta_\mathrm{int}^\mathrm{max}=72\%$ which corresponds to a maximum total device efficiency of $\eta_\mathrm{dev}^\mathrm{max}\approx 10\%$ with $\eta_\mathrm{dev}=\eta_\mathrm{int}\eta_\mathrm{loss}$, with $\eta_\mathrm{loss}= 14\%$. The SNR, defined as the background subtracted conversion signal over the background, follows the expected behavior (blue line) showing a drop for low pump powers due to the dark count limitation of our detector ($DC_{1552}=10\,\mathrm{Hz}$) as well as a decrease for very high pump powers due to the non-linear dependence of $\eta_\mathrm{int}$ on $P_\mathrm{pump}$. For comparison we also included a trace of the SNR measured without the etalon (grey squares) which shows significantly worse filtering. The inset shows the SNR depending on the mean input photon number $\mu_\mathrm{in}$ for full filtering (including the etalon) for a fixed pump power of $P_\mathrm{pump}=287\,\mathrm{mW}$. We observe the expected linear dependence $\mathrm{SNR}=\mathrm{SNR}_\mathrm{max}\times\mu_\mathrm{in}$ with $\mathrm{SNR}_\mathrm{max}=452$ for a single photon input (i.e. $\mu_\mathrm{in}=1$). This represents a more than fivefold improvement compared to former reported results \cite{Albrecht2014}.

\subsection{Combined QM and QFCD}
\label{sec:Combined}
\begin{figure}[t]
\centering
\includegraphics[width=\linewidth]{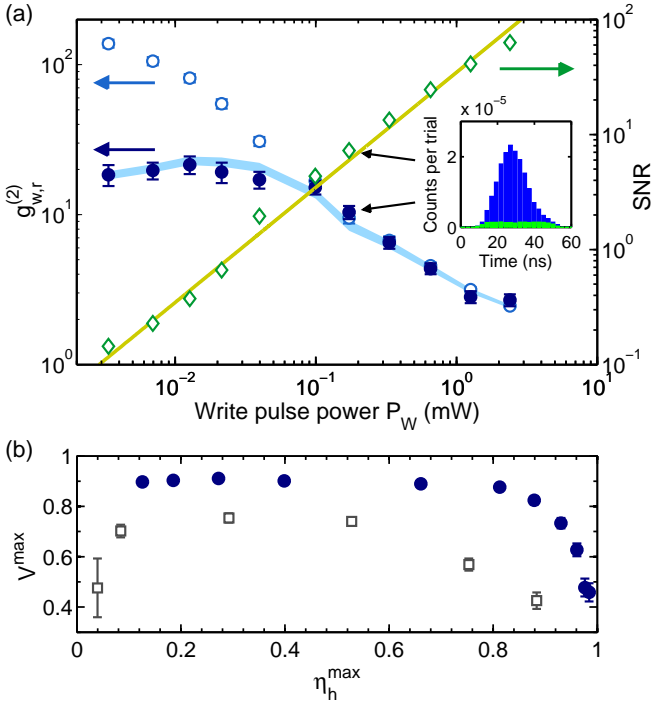}
\caption{(color online) (a) Normalized cross-correlation of the unconverted write photons (blue circles, left axis) and the converted ones (blue dots) with the read photons and SNR of the converted write photons (green diamonds, right axis, errorbars smaller than symbol size) vs. peak power of the write pulse. The blue shaded area corresponds to the expected $g^{(2)}_{cw,r}$ as inferred from Eq.~(\ref{g2cwr}), and the SNR is fitted by a linear regression (green line). The inset shows as an example the detected shape of the converted write photon for $P_W=333\,\mu\mathrm{W}$. The pump power was fixed at $P_\mathrm{pump}=290\,\mathrm{mW}$. (b) Extrapolated maximal visibility $V^\mathrm{max}$ for interfering two frequency converted photons from two atomic ensembles depending on the maximal heralding efficiency $\eta_h^\mathrm{max}$ (blue dots for full filtering, grey squares without etalon).}
\label{Figure3}
\end{figure}

Next, we combined the QFCD with the cold atomic QM to convert the write photons from $780\,\mathrm{nm}$ to $1552\,\mathrm{nm}$ and investigate the joint properties of the telecom photons and the atomic spin-wave stored in the QM. To create the spin-wave weak Gaussian shaped write pulses of $\tau_W=20\,\mathrm{ns}$ duration were sent and the width of write photon detection gate was set to $40\,\mathrm{ns}$ (cf. green area in the inset of Fig.~\ref{Figure3}(a)). To gain information about the spin-wave, we sent a read pulse ($\tau_R=35\,\mathrm{ns}$, $P_R=190\,\mu\mathrm{W}$) resonant to the $|s\rangle\leftrightarrow|e\rangle$ transition to convert the spin-wave back into a single read photon. Due to collective interference of the atoms, the read photon is emitted in a spatial mode given by the phase matching condition $\textbf{k}_r = \textbf{k}_R + \textbf{k}_W - \textbf{k}_w$, with $\textbf{k}_{r,R,W,w}$ denoting the respective wave vectors of the single photons and pulses \cite{Duan2001}. The read photon is then polarization filtered before being sent through a monolithic Fabry-Perot cavity ($\eta_{cav}\approx 20\%$ total transmission, including cavity transmission and subsequent fiber coupling) for spectral filtering and finally detected in a window of $100\,\mathrm{ns}$ by a silicon SPD (Excelitas SPCM-AQRH-14) with $\eta_{d,780}=40\%$ efficiency. The retrieval efficiency is defined as the probability to map a heralded spin-wave onto a read photon. Its raw value is calculated as $\eta_{ret}=p_{cw,r}/p_{cw}$, where $p_{cw,r}$ is the probability per trial to detect a coincidence between a converted write and a read photon and $p_{cw}$ is the probability per trial to obtain a detection event in the converted write photon detector. The fiber-coupled retrieval efficiency $\eta_{ret}^{fiber}=\eta_{ret}/(\eta_{cav}\eta_{d,780})$ corresponds to the probability of finding a read photon in the optical fiber after the vacuum cell, i.e. corrected for filtering and detector efficiency only.

To demonstrate that the conversion of the write photon preserves its quantum character, we measured the normalized second-order cross-correlation between the converted write photon and the read photon defined as $g^{(2)}_{cw,r}=p_{cw,r}/(p_{cw}p_{r})$. For comparison we also took the cross-correlation $g^{(2)}_{w,r}$ without write photon conversion, for which we replaced the QFCD by a Fabry-Perot filtering cavity with similar characteristics as the one used for the read photons but resonant with the write photons. The obtained data are shown in the Fig.~\ref{Figure3}(a) as blue dots for $g^{(2)}_{cw,r}$ and blue circles for $g^{(2)}_{w,r}$ vs. the applied power of the write pulse. We observe the highest cross-correlation of $g^{(2)}_{cw,r}\approx 20$ for a write pulse power of $P_W\approx 10\,\mu W$. For higher $P_W$, $g^{(2)}_{cw,r}$ decreases, as expected for a DLCZ-type QM. For lower values of $P_W$, $g^{(2)}_{cw,r}$ slightly drops due to noise introduced by the QFCD and the dark counts of the SPDs \cite{Sekatski2012}. This also explains the deviation of $g^{(2)}_{cw,r}$ from $g^{(2)}_{w,r}$ in the low $P_W$ regime. The measured  $g^{(2)}_{cw,r}$ in  Fig.~\ref{Figure3}(a) are well above the classical limit of 2, assuming thermal statistics for the write and read beams (see below). This shows that we can operate the combined QM-QFC device for a large range of write pulse powers in a highly non-classical regime. The experimental data follow well the expected behavior taking into account the background noise created by the QFCD pump laser (indicated by the blue shaded area) which can be deduced from
\begin{equation} \label{g2cwr}
g^{(2)}_{cw,r} = \frac{g^{(2)}_{w,r}+\mathrm{SNR}^{-1}}{1+\mathrm{SNR}^{-1}}.
\end{equation}
Here, $g^{(2)}_{w,r}$ denotes the measured cross-correlation if the write photon is sent through a filtering cavity (similar to the read photon cavity) instead of the QFCD and SNR is the signal to noise ratio of the converted write photon. $\mathrm{SNR}=(p_{cw}-p_N)/p_N$, where $p_N$ is the probability to have a detection when the write photon is blocked before the QFCD (see also Supplemental Material). The good agreement between the experimental data and the simple model suggests that the noise generated by the QFCD pump beam is the main limiting factor for the value of $g^{(2)}_{cw,r}$. 

Moreover, we proved unambigusously the high degree non-classical correlations between the converted write photons and the retrieved read photons by violating the Cauchy-Schwarz inequality for classical light, given by
\begin{equation} \label{CSI}
R = \frac{{(g^{(2)}_{cw,r})}^2}{g^{(2)}_{cw,cw} \cdot g^{(2)}_{r,r}} \le 1
\end{equation}
where where $g^{(2)}_{cw,cw}=\frac{p_{cw,cw}}{p_{cw} p_{cw}}$ ($g^{(2)}_{r,r}=\frac{p_{r,r}}{p_r p_r}$) denotes the unheralded auto-correlation function of the converted write (read) photons. The measured correlation values for different write powers and the inferred Cauchy-Schwarz parameter $R$ are given in Table~\ref{table1}. Even for relatively high write pulse powers we clearly violate equation~(\ref{CSI}). For $P_W=0.17\,\mathrm{mW}$ we obtain $R=31$ violating the Cauchy-Schwarz inequality by more than four standard deviations, clearly demonstrating strong non-classical correlations between the converted write photons and the retrieved read photons.

\begin{table}
\caption{Measured values of the coincidence detection probability $p_{cw,r}$, the cross-correlation $g^{(2)}_{cw,r}$ and the unheralded auto-correlations $g^{(2)}_{cw,cw}$ and $g^{(2)}_{r,r}$ of the converted write photons and read photons for different write pulse powers $P_W$. Errors correspond to $\pm1$ standard deviation. The Cauchy-Schwarz parameter $R$ is calculated from Eq.~(\ref{CSI}).}
\centering
\begin{tabular}{ K{1.5cm} | K{1.5cm} | K{1.1cm} K{1.1cm} K{1.1cm} || K{1.1cm} }
$P_W\,\mathrm{[mW]}$	&$p_{cw,r}\,[\%]$			& $g^{(2)}_{cw,r}$ 	& $g^{(2)}_{cw,cw}$ 	& $g^{(2)}_{r,r}$ 	& $R$ \\ 
\hline
2.39 					& $4.2\cdot 10^{-3}$		& $2.48(6)$				& $2.0(2)$				& $2.16(9)$			& $1.4(2)$ \\
0.65 					& $1.2\cdot 10^{-3}$		& $4.49(8)$				& $2.3(3)$				& $2.04(9)$ 		& $4.4(7)$ \\
0.17 					& $0.3\cdot 10^{-3}$		& $9.9(2)$				& $1.6(4)$ 				& $2.0(1)$ 			& $31(7)$ \\
\end{tabular}
\label{table1}
\end{table}

In addition to non-classical correlations, another requirement to build a reliable QR is to achieve a high SNR in the detection of the converted heralding photon. Hence, we investigated the SNR of the frequency converted write photon depending on the write pulse power. The results are shown as green diamonds in Fig.~\ref{Figure3}(a). We observe the expected linear increase of the SNR with write pulse power as $\mathrm{SNR}\propto P_W$. For large $P_W$ we observe $\mathrm{SNRs} > 50$ while still being in the the non-classical regime. 

From the data presented in Fig.~\ref{Figure3}(a), we can obtain more insight about the performance of our combined QM-QFC device for quantum information protocols. From the measured $g^{(2)}_{cw,r}$, we can infer the maximal visibility $V^{max}$ that could be achieved in a two photon interference experiment. For example if the atomic qubit was entangled with a converted photonic qubit (e.g. in polarization \cite{DeRiedmatten2006} or time-bin \cite{Albrecht2015}) or in a two ensemble entanglement experiment \cite{Laurat2007} we can infer $V^{max}=(g^{(2)}_{cw,r}-1)/(g^{(2)}_{cw,r}+1)$. Here, it is assumed that the QFCD is phase-preserving \cite{Fernandez-Gonzalvo2013,Ikuta2013}. Also from the measured SNR, we can infer the maximal heralding efficiency $\eta_h^{max}=\mathrm{SNR}/(\mathrm{SNR}+1)$ of the combined QM-QFC device for heralding the presence of a spin-wave in the ensemble. The infered $V^{max}$ is plotted as a function of $\eta_h^{max}$ in Fig.~\ref{Figure3}(b). We infer a visibility of $V\approx 90\%$ up to $80\%$ heralding efficiency for full filtering (blue dots).  $V^{max}$ decreases for higher $\eta_h^{max}$, but we still obtain $V^{max}> 1/\sqrt{2}$, potentially enabling a violation of the  Clauser-Horne-Shimony-Holt (CHSH) type Bell inequality, for $\eta_h^{max}> 90 \%$.  This confirms the suitability of the combined QM-QFC device as a elemental building block of a QR for long distance quantum communication. If the etalon is removed, $V^{max}$ drops significantly (cf. grey squares) clearly demonstrating the importance of ultra-narrowband filtering for the all-solid state based QFC approach. We stress that these values are given here only to infer the potential of our device for quantum information experiments, and should be confirmed with further measurements.

\begin{figure}
\centering
\includegraphics[width=\linewidth]{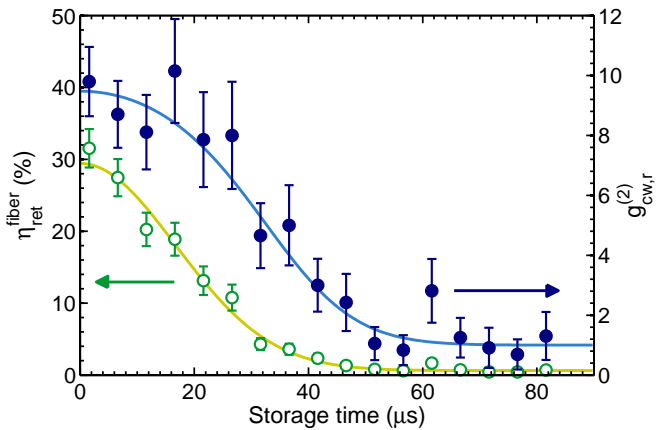}
\caption{(color online) Retrieval efficiency (green circles, left axis) and normalized cross-correlation of the converted write photons and read photons (blue dots, right axis) vs. storage time in the QM. The write and pump powers were fixed at $P_W=0.18\,\mathrm{mW}$ and $P_\mathrm{pump}=290\,\mathrm{mW}$.}
\label{Figure4}
\end{figure} 

Finally, we investigated the capability of the combined QM-QFC device to preserve the non-classical correlations of the converted write photons and the stored spin-wave depending on the storage time in the QM. The retrieval efficiency $\eta_{ret}$ (green circles in Fig.~\ref{Figure4}) decreases over storage time due to dephasing of the stored spin-wave mainly induced by thermal atomic motion and to a smaller degree by external spurious magnetic field gradients. The decay can be fitted with a theoretical model introduced in the Supplemental Material (see green line in Fig.~\ref{Figure4}) giving a decay time of $\tau=23.6\pm0.8\,\mu\mathrm{s}$. However, the storage time is not a fundamental limitation here, as it could be increased by orders of magnitude using other techniques \cite{Radnaev2010,Dudin2010,Bao2012,Yang2016,Jiang2016}. The normalized cross-correlation $g^{(2)}_{cw,r}$ between the converted write photons and the retrieved read photons is shown as blue dots in Fig.~\ref{Figure4} for a write pulse power of $P_W=0.18\,\mathrm{mW}$. We observe the expected decay of $g^{(2)}_{cw,r}$ and fit the data with the above mentioned model, giving a decay time of $\tau=25.8\pm1.2\,\mu\mathrm{s}$ which is consistent with the result obtained when fitting $\eta_{ret}$. Fig.~\ref{Figure4} shows that we stay in the non-classical regime ($g^{(2)}_{cw,r}>2$) up to storage times of about $40\,\mu\mathrm{s}$ which corresponds to a fiber transmission length of approximately $8\,\mathrm{km}$.

\subsection{Discussion}
\label{sec:Discussion}

The performance of the QFCD is currently mainly limited by technical issues like coupling efficiencies in the PPLN wave-guide and into the final optical fiber as well as transmission efficiencies through the filtering stage and other optical surfaces. However, with the current device conversion efficiency of about $10\,\%$ and typical fiber-absorptions of $0.2\,\mathrm{dB/km}$ for $1552\,\mathrm{nm}$ and $3.5\,\mathrm{dB/km}$ for $780\,\mathrm{nm}$, the unconverted photon traveling in a $780\,\mathrm{nm}$ fiber would experience a higher loss after around $3\,\mathrm{km}$ than a frequency converted photon traveling in a telecom fiber. Hence, even with this seemingly low conversion efficiency, QFC beats direct transmission already after a few km.

Second, we note that a QFCD with a given device efficiency $\eta_\mathrm{dev}$ is equivalent in terms of loss to an additional fiber length of $L=-10/0.2 \cdot \log(\eta_\mathrm{dev})$. For the current device efficiency of $10\,\%$, this corresponds to an additional loss of $-10\,\mathrm{dB}$ meaning an equivalent of $50\,\mathrm{km}$ extra fiber in each arm of a telecom quantum repeater. A device efficiency of $50\,\%$ would corresponds to $15\,\mathrm{km}$ of additional fiber in each arm.

Finally, we note that to alleviate the requirements for spectral filtering and thus to increase the QFC efficiency, also different conversion strategies with further separated wavelengths could be considered \cite{Pelc2012}. A larger wavelength separation would decrease the Raman noise or could even suppress it completely. However, to convert the $780\,\mathrm{nm}$ photon into the telecom C-band ($1530\,\mathrm{nm}-1565\,\mathrm{nm}$) where the losses in optical fibers are the lowest, not much flexibility is possible. The Raman noise is present up to $700\,\mathrm{cm}^{-1}$ ($21\,\mathrm{THz}$) away from the excitation pump, as measured in \cite{Zaske2011}. In our case, the frequency separation between the pump at $1569\,\mathrm{nm}$ and the target wavelength at $1552\,\mathrm{nm}$ is $2.1\,\mathrm{THz}$. Using a pump at the edge of the gain spectrum of Erbium amplifiers, around $1605\,\mathrm{nm}$, the separation in frequency between that pump and the target wavelength at $1517,\mathrm{nm}$ would be $11\,\mathrm{THz}$, which is still inside the Raman noise window. The solution for a noise-free conversion as mentioned in Ref.~\cite{Pelc2012} would indeed be to use a pump at around $2000\,\mathrm{nm}$ converting the $780\,\mathrm{nm}$ photons to $1280\,\mathrm{nm}$ into the telecom O-band (frequency difference of $84\,\mathrm{THz}$). This however, would come with the expenses of higher transmission losses in fibers, a more problematic mode matching in the non-linear wave-guide, and the need for more sophisticated technical resources.

\section{Conclusion}
\label{sec:Conclusion}

In conclusion, we demonstrated highly non-classical correlations between a frequency converted telecom C-band photon and a spin-wave stored in an atomic quantum memory. The photon heralding the spin-wave was converted from 780 nm to 1552 nm using an integrated non-linear waveguide.  We showed that by improved optical filtering very high signal to noise ratios up to $\mathrm{SNR}_\mathrm{max}=452$ for a weak coherent input pulse with in average one photon per pulse could be achieved. This was the key to obtain high non-classical correlations between the converted write and read photons up to $g^{(2)}_{w,r}\approx 20$ when the QFCD was combined with the QM, as well as high SNRs for the detection of the converted write photon, leading to high spin-wave heralding efficiencies. Moreover, we proved that the quantum character of the converted write photons and read photons are preserved by violating the Cauchy-Schwarz inequality by more than four standard deviations. Finally, we demonstrated that the non-classical correlations between the heralding telecom write photon and the near infrared read photon could be stored in the QM up to $40\,\mu\mathrm{s}$. Our experiment shows that quantum frequency conversion based on integrated non-linear waveguide is a viable approach to create quantum correlation between telecom photons and long-lived spin-waves. This approach offers significant advantages in terms of wavelength flexibility, robustness and simplicity compared to frequency conversion processes in cold atomic gases. The robust all-solid-state quantum frequency conversion of the heralding write photons in combination with the preservation and  storage of non-classical correlations in a DLCZ-QM is an important step towards the interconnection of matter nodes operating at different wavelengths and the realistic implementation of an elementary telecom quantum repeater link.

\section*{Acknowledgments}
We thank Xavier Fernandez-Gonzalvo and Matteo Cristiani for helpful discussions and support during the initial setup of the experiment. We acknowledge financial support by the ERC starting grant QuLIMA, by the Spanish Ministry of Economy and Competitiveness (MINECO) and the Fondo Europeo de Desarrollo Regional (FEDER) through grant FIS2012-37569, by MINECO Severo Ochoa through grant SEV-2015-0522 and by AGAUR via 2014 SGR 1554 and the Fundaci\'{o} Privada Cellex. P.F. acknowledges the International PhD-fellowship program "la Caixa"-Severo Ochoa @ ICFO. G.H. acknowledges support by the ICFOnest+ international postdoctoral fellowship program.


\begin{thebibliography}{45}%
\makeatletter
\providecommand \@ifxundefined [1]{%
 \@ifx{#1\undefined}
}%
\providecommand \@ifnum [1]{%
 \ifnum #1\expandafter \@firstoftwo
 \else \expandafter \@secondoftwo
 \fi
}%
\providecommand \@ifx [1]{%
 \ifx #1\expandafter \@firstoftwo
 \else \expandafter \@secondoftwo
 \fi
}%
\providecommand \natexlab [1]{#1}%
\providecommand \enquote  [1]{``#1''}%
\providecommand \bibnamefont  [1]{#1}%
\providecommand \bibfnamefont [1]{#1}%
\providecommand \citenamefont [1]{#1}%
\providecommand \href@noop [0]{\@secondoftwo}%
\providecommand \href [0]{\begingroup \@sanitize@url \@href}%
\providecommand \@href[1]{\@@startlink{#1}\@@href}%
\providecommand \@@href[1]{\endgroup#1\@@endlink}%
\providecommand \@sanitize@url [0]{\catcode `\\12\catcode `\$12\catcode
  `\&12\catcode `\#12\catcode `\^12\catcode `\_12\catcode `\%12\relax}%
\providecommand \@@startlink[1]{}%
\providecommand \@@endlink[0]{}%
\providecommand \url  [0]{\begingroup\@sanitize@url \@url }%
\providecommand \@url [1]{\endgroup\@href {#1}{\urlprefix }}%
\providecommand \urlprefix  [0]{URL }%
\providecommand \Eprint [0]{\href }%
\providecommand \doibase [0]{http://dx.doi.org/}%
\providecommand \selectlanguage [0]{\@gobble}%
\providecommand \bibinfo  [0]{\@secondoftwo}%
\providecommand \bibfield  [0]{\@secondoftwo}%
\providecommand \translation [1]{[#1]}%
\providecommand \BibitemOpen [0]{}%
\providecommand \bibitemStop [0]{}%
\providecommand \bibitemNoStop [0]{.\EOS\space}%
\providecommand \EOS [0]{\spacefactor3000\relax}%
\providecommand \BibitemShut  [1]{\csname bibitem#1\endcsname}%
\let\auto@bib@innerbib\@empty
\bibitem [{\citenamefont {Korzh}\ \emph {et~al.}(2015)\citenamefont {Korzh},
  \citenamefont {Lim}, \citenamefont {Houlmann}, \citenamefont {Gisin},
  \citenamefont {Li}, \citenamefont {Nolan}, \citenamefont {Sanguinetti},
  \citenamefont {Thew},\ and\ \citenamefont {Zbinden}}]{Korzh2015}%
  \BibitemOpen
  \bibfield  {author} {\bibinfo {author} {\bibfnamefont {B.}~\bibnamefont
  {Korzh}}, \bibinfo {author} {\bibfnamefont {C.~C.~W.}\ \bibnamefont {Lim}},
  \bibinfo {author} {\bibfnamefont {R.}~\bibnamefont {Houlmann}}, \bibinfo
  {author} {\bibfnamefont {N.}~\bibnamefont {Gisin}}, \bibinfo {author}
  {\bibfnamefont {M.~J.}\ \bibnamefont {Li}}, \bibinfo {author} {\bibfnamefont
  {D.}~\bibnamefont {Nolan}}, \bibinfo {author} {\bibfnamefont
  {B.}~\bibnamefont {Sanguinetti}}, \bibinfo {author} {\bibfnamefont
  {R.}~\bibnamefont {Thew}}, \ and\ \bibinfo {author} {\bibfnamefont
  {H.}~\bibnamefont {Zbinden}},\ }\href {\doibase 10.1038/nphoton.2014.327}
  {\bibfield  {journal} {\bibinfo  {journal} {Nature Photonics}\ }\textbf
  {\bibinfo {volume} {9}},\ \bibinfo {pages} {163} (\bibinfo {year}
  {2015})}\BibitemShut {NoStop}%
\bibitem [{\citenamefont {Briegel}\ \emph {et~al.}(1998)\citenamefont
  {Briegel}, \citenamefont {D{\"{u}}r}, \citenamefont {Cirac},\ and\
  \citenamefont {Zoller}}]{Briegel1998}%
  \BibitemOpen
  \bibfield  {author} {\bibinfo {author} {\bibfnamefont {H.-J.}\ \bibnamefont
  {Briegel}}, \bibinfo {author} {\bibfnamefont {W.}~\bibnamefont {D{\"{u}}r}},
  \bibinfo {author} {\bibfnamefont {J.~I.}\ \bibnamefont {Cirac}}, \ and\
  \bibinfo {author} {\bibfnamefont {P.}~\bibnamefont {Zoller}},\ }\href
  {\doibase 10.1103/PhysRevLett.81.5932} {\bibfield  {journal} {\bibinfo
  {journal} {Physical Review Letters}\ }\textbf {\bibinfo {volume} {81}},\
  \bibinfo {pages} {5932} (\bibinfo {year} {1998})}\BibitemShut {NoStop}%
\bibitem [{\citenamefont {Sangouard}\ \emph {et~al.}(2011)\citenamefont
  {Sangouard}, \citenamefont {Simon}, \citenamefont {de~Riedmatten},\ and\
  \citenamefont {Gisin}}]{Sangouard2011}%
  \BibitemOpen
  \bibfield  {author} {\bibinfo {author} {\bibfnamefont {N.}~\bibnamefont
  {Sangouard}}, \bibinfo {author} {\bibfnamefont {C.}~\bibnamefont {Simon}},
  \bibinfo {author} {\bibfnamefont {H.}~\bibnamefont {de~Riedmatten}}, \ and\
  \bibinfo {author} {\bibfnamefont {N.}~\bibnamefont {Gisin}},\ }\href
  {\doibase 10.1103/RevModPhys.83.33} {\bibfield  {journal} {\bibinfo
  {journal} {Review of Modern Physics}\ }\textbf {\bibinfo {volume} {83}},\
  \bibinfo {pages} {33} (\bibinfo {year} {2011})}\BibitemShut {NoStop}%
\bibitem [{\citenamefont {Afzelius}\ \emph {et~al.}(2015)\citenamefont
  {Afzelius}, \citenamefont {Gisin},\ and\ \citenamefont
  {de~Riedmatten}}]{Afzelius2015}%
  \BibitemOpen
  \bibfield  {author} {\bibinfo {author} {\bibfnamefont {M.}~\bibnamefont
  {Afzelius}}, \bibinfo {author} {\bibfnamefont {N.}~\bibnamefont {Gisin}}, \
  and\ \bibinfo {author} {\bibfnamefont {H.}~\bibnamefont {de~Riedmatten}},\
  }\href {\doibase 10.1063/PT.3.3021} {\bibfield  {journal} {\bibinfo
  {journal} {Physics Today}\ }\textbf {\bibinfo {volume} {68}},\ \bibinfo
  {pages} {42} (\bibinfo {year} {2015})}\BibitemShut {NoStop}%
\bibitem [{\citenamefont {Lvovsky}\ \emph {et~al.}(2009)\citenamefont
  {Lvovsky}, \citenamefont {Sanders},\ and\ \citenamefont
  {Tittel}}]{Lvovsky2009}%
  \BibitemOpen
  \bibfield  {author} {\bibinfo {author} {\bibfnamefont {A.~I.}\ \bibnamefont
  {Lvovsky}}, \bibinfo {author} {\bibfnamefont {B.~C.}\ \bibnamefont
  {Sanders}}, \ and\ \bibinfo {author} {\bibfnamefont {W.}~\bibnamefont
  {Tittel}},\ }\href {\doibase 10.1038/nphoton.2009.231} {\bibfield  {journal}
  {\bibinfo  {journal} {Nature Photonics}\ }\textbf {\bibinfo {volume} {3}},\
  \bibinfo {pages} {706} (\bibinfo {year} {2009})}\BibitemShut {NoStop}%
\bibitem [{\citenamefont {Hammerer}\ \emph {et~al.}(2010)\citenamefont
  {Hammerer}, \citenamefont {S{\o}rensen},\ and\ \citenamefont
  {Polzik}}]{Hammerer2010}%
  \BibitemOpen
  \bibfield  {author} {\bibinfo {author} {\bibfnamefont {K.}~\bibnamefont
  {Hammerer}}, \bibinfo {author} {\bibfnamefont {A.~S.}\ \bibnamefont
  {S{\o}rensen}}, \ and\ \bibinfo {author} {\bibfnamefont {E.~S.}\ \bibnamefont
  {Polzik}},\ }\href {\doibase 10.1103/RevModPhys.82.1041} {\bibfield
  {journal} {\bibinfo  {journal} {Reviews of Modern Physics}\ }\textbf
  {\bibinfo {volume} {82}},\ \bibinfo {pages} {1041} (\bibinfo {year}
  {2010})}\BibitemShut {NoStop}%
\bibitem [{\citenamefont {Simon}\ \emph {et~al.}(2010)\citenamefont {Simon},
  \citenamefont {Afzelius}, \citenamefont {Appel}, \citenamefont {{Boyer de la
  Giroday}}, \citenamefont {Dewhurst}, \citenamefont {Gisin}, \citenamefont
  {Hu}, \citenamefont {Jelezko}, \citenamefont {Kr{\"{o}}ll}, \citenamefont
  {M{\"{u}}ller}, \citenamefont {Nunn}, \citenamefont {Polzik}, \citenamefont
  {Rarity}, \citenamefont {{De Riedmatten}}, \citenamefont {Rosenfeld},
  \citenamefont {Shields}, \citenamefont {Sk{\"{o}}ld}, \citenamefont
  {Stevenson}, \citenamefont {Thew}, \citenamefont {Walmsley}, \citenamefont
  {Weber}, \citenamefont {Weinfurter}, \citenamefont {Wrachtrup},\ and\
  \citenamefont {Young}}]{Simon2010}%
  \BibitemOpen
  \bibfield  {author} {\bibinfo {author} {\bibfnamefont {C.}~\bibnamefont
  {Simon}}, \bibinfo {author} {\bibfnamefont {M.}~\bibnamefont {Afzelius}},
  \bibinfo {author} {\bibfnamefont {J.}~\bibnamefont {Appel}}, \bibinfo
  {author} {\bibfnamefont {A.}~\bibnamefont {{Boyer de la Giroday}}}, \bibinfo
  {author} {\bibfnamefont {S.~J.}\ \bibnamefont {Dewhurst}}, \bibinfo {author}
  {\bibfnamefont {N.}~\bibnamefont {Gisin}}, \bibinfo {author} {\bibfnamefont
  {C.~Y.}\ \bibnamefont {Hu}}, \bibinfo {author} {\bibfnamefont
  {F.}~\bibnamefont {Jelezko}}, \bibinfo {author} {\bibfnamefont
  {S.}~\bibnamefont {Kr{\"{o}}ll}}, \bibinfo {author} {\bibfnamefont {J.~H.}\
  \bibnamefont {M{\"{u}}ller}}, \bibinfo {author} {\bibfnamefont
  {J.}~\bibnamefont {Nunn}}, \bibinfo {author} {\bibfnamefont {E.~S.}\
  \bibnamefont {Polzik}}, \bibinfo {author} {\bibfnamefont {J.~G.}\
  \bibnamefont {Rarity}}, \bibinfo {author} {\bibfnamefont {H.}~\bibnamefont
  {{De Riedmatten}}}, \bibinfo {author} {\bibfnamefont {W.}~\bibnamefont
  {Rosenfeld}}, \bibinfo {author} {\bibfnamefont {A.~J.}\ \bibnamefont
  {Shields}}, \bibinfo {author} {\bibfnamefont {N.}~\bibnamefont
  {Sk{\"{o}}ld}}, \bibinfo {author} {\bibfnamefont {R.~M.}\ \bibnamefont
  {Stevenson}}, \bibinfo {author} {\bibfnamefont {R.}~\bibnamefont {Thew}},
  \bibinfo {author} {\bibfnamefont {I.~A.}\ \bibnamefont {Walmsley}}, \bibinfo
  {author} {\bibfnamefont {M.~C.}\ \bibnamefont {Weber}}, \bibinfo {author}
  {\bibfnamefont {H.}~\bibnamefont {Weinfurter}}, \bibinfo {author}
  {\bibfnamefont {J.}~\bibnamefont {Wrachtrup}}, \ and\ \bibinfo {author}
  {\bibfnamefont {R.~J.}\ \bibnamefont {Young}},\ }\href {\doibase
  10.1140/epjd/e2010-00103-y} {\bibfield  {journal} {\bibinfo  {journal} {The
  European Physical Journal D}\ }\textbf {\bibinfo {volume} {58}},\ \bibinfo
  {pages} {1} (\bibinfo {year} {2010})}\BibitemShut {NoStop}%
\bibitem [{\citenamefont {Bussi{\`{e}}res}\ \emph {et~al.}(2013)\citenamefont
  {Bussi{\`{e}}res}, \citenamefont {Sangouard}, \citenamefont {Afzelius},
  \citenamefont {de~Riedmatten}, \citenamefont {Simon},\ and\ \citenamefont
  {Tittel}}]{Bussieres2013}%
  \BibitemOpen
  \bibfield  {author} {\bibinfo {author} {\bibfnamefont {F.}~\bibnamefont
  {Bussi{\`{e}}res}}, \bibinfo {author} {\bibfnamefont {N.}~\bibnamefont
  {Sangouard}}, \bibinfo {author} {\bibfnamefont {M.}~\bibnamefont {Afzelius}},
  \bibinfo {author} {\bibfnamefont {H.}~\bibnamefont {de~Riedmatten}}, \bibinfo
  {author} {\bibfnamefont {C.}~\bibnamefont {Simon}}, \ and\ \bibinfo {author}
  {\bibfnamefont {W.}~\bibnamefont {Tittel}},\ }\href {\doibase
  10.1080/09500340.2013.856482} {\bibfield  {journal} {\bibinfo  {journal}
  {Journal of Modern Optics}\ }\textbf {\bibinfo {volume} {60}},\ \bibinfo
  {pages} {1519} (\bibinfo {year} {2013})}\BibitemShut {NoStop}%
\bibitem [{\citenamefont {Chou}\ \emph {et~al.}(2005)\citenamefont {Chou},
  \citenamefont {de~Riedmatten}, \citenamefont {Felinto}, \citenamefont
  {Polyakov}, \citenamefont {van Enk},\ and\ \citenamefont
  {Kimble}}]{Chou2005}%
  \BibitemOpen
  \bibfield  {author} {\bibinfo {author} {\bibfnamefont {C.~W.}\ \bibnamefont
  {Chou}}, \bibinfo {author} {\bibfnamefont {H.}~\bibnamefont {de~Riedmatten}},
  \bibinfo {author} {\bibfnamefont {D.}~\bibnamefont {Felinto}}, \bibinfo
  {author} {\bibfnamefont {S.~V.}\ \bibnamefont {Polyakov}}, \bibinfo {author}
  {\bibfnamefont {S.~J.}\ \bibnamefont {van Enk}}, \ and\ \bibinfo {author}
  {\bibfnamefont {H.~J.}\ \bibnamefont {Kimble}},\ }\href {\doibase
  10.1038/nature04353} {\bibfield  {journal} {\bibinfo  {journal} {Nature}\
  }\textbf {\bibinfo {volume} {438}},\ \bibinfo {pages} {828} (\bibinfo {year}
  {2005})}\BibitemShut {NoStop}%
\bibitem [{\citenamefont {Chou}\ \emph {et~al.}(2007)\citenamefont {Chou},
  \citenamefont {Laurat}, \citenamefont {Deng}, \citenamefont {Choi},
  \citenamefont {de~Riedmatten}, \citenamefont {Felinto},\ and\ \citenamefont
  {Kimble}}]{Chou2007}%
  \BibitemOpen
  \bibfield  {author} {\bibinfo {author} {\bibfnamefont {C.-W.}\ \bibnamefont
  {Chou}}, \bibinfo {author} {\bibfnamefont {J.}~\bibnamefont {Laurat}},
  \bibinfo {author} {\bibfnamefont {H.}~\bibnamefont {Deng}}, \bibinfo {author}
  {\bibfnamefont {K.~S.}\ \bibnamefont {Choi}}, \bibinfo {author}
  {\bibfnamefont {H.}~\bibnamefont {de~Riedmatten}}, \bibinfo {author}
  {\bibfnamefont {D.}~\bibnamefont {Felinto}}, \ and\ \bibinfo {author}
  {\bibfnamefont {H.~J.}\ \bibnamefont {Kimble}},\ }\href {\doibase
  10.1126/science.1140300} {\bibfield  {journal} {\bibinfo  {journal} {Science
  (New York, N.Y.)}\ }\textbf {\bibinfo {volume} {316}},\ \bibinfo {pages}
  {1316} (\bibinfo {year} {2007})}\BibitemShut {NoStop}%
\bibitem [{\citenamefont {Yuan}\ \emph {et~al.}(2008)\citenamefont {Yuan},
  \citenamefont {Chen}, \citenamefont {Zhao}, \citenamefont {Chen},
  \citenamefont {Schmiedmayer},\ and\ \citenamefont {Pan}}]{Yuan2008}%
  \BibitemOpen
  \bibfield  {author} {\bibinfo {author} {\bibfnamefont {Z.-S.}\ \bibnamefont
  {Yuan}}, \bibinfo {author} {\bibfnamefont {Y.-A.}\ \bibnamefont {Chen}},
  \bibinfo {author} {\bibfnamefont {B.}~\bibnamefont {Zhao}}, \bibinfo {author}
  {\bibfnamefont {S.}~\bibnamefont {Chen}}, \bibinfo {author} {\bibfnamefont
  {J.}~\bibnamefont {Schmiedmayer}}, \ and\ \bibinfo {author} {\bibfnamefont
  {J.-W.}\ \bibnamefont {Pan}},\ }\href {\doibase 10.1038/nature07241}
  {\bibfield  {journal} {\bibinfo  {journal} {Nature}\ }\textbf {\bibinfo
  {volume} {454}},\ \bibinfo {pages} {1098} (\bibinfo {year}
  {2008})}\BibitemShut {NoStop}%
\bibitem [{\citenamefont {Moehring}\ \emph {et~al.}(2007)\citenamefont
  {Moehring}, \citenamefont {Maunz}, \citenamefont {Olmschenk}, \citenamefont
  {Younge}, \citenamefont {Matsukevich}, \citenamefont {Duan},\ and\
  \citenamefont {Monroe}}]{Moehring2007}%
  \BibitemOpen
  \bibfield  {author} {\bibinfo {author} {\bibfnamefont {D.~L.}\ \bibnamefont
  {Moehring}}, \bibinfo {author} {\bibfnamefont {P.}~\bibnamefont {Maunz}},
  \bibinfo {author} {\bibfnamefont {S.}~\bibnamefont {Olmschenk}}, \bibinfo
  {author} {\bibfnamefont {K.~C.}\ \bibnamefont {Younge}}, \bibinfo {author}
  {\bibfnamefont {D.~N.}\ \bibnamefont {Matsukevich}}, \bibinfo {author}
  {\bibfnamefont {L.-M.}\ \bibnamefont {Duan}}, \ and\ \bibinfo {author}
  {\bibfnamefont {C.}~\bibnamefont {Monroe}},\ }\href {\doibase
  10.1038/nature06118} {\bibfield  {journal} {\bibinfo  {journal} {Nature}\
  }\textbf {\bibinfo {volume} {449}},\ \bibinfo {pages} {68} (\bibinfo {year}
  {2007})}\BibitemShut {NoStop}%
\bibitem [{\citenamefont {Hensen}\ \emph {et~al.}(2015)\citenamefont {Hensen},
  \citenamefont {Bernien}, \citenamefont {Dr{\'{e}}au}, \citenamefont
  {Reiserer}, \citenamefont {Kalb}, \citenamefont {Blok}, \citenamefont
  {Ruitenberg}, \citenamefont {Vermeulen}, \citenamefont {Schouten},
  \citenamefont {Abell{\'{a}}n}, \citenamefont {Amaya}, \citenamefont
  {Pruneri}, \citenamefont {Mitchell}, \citenamefont {Markham}, \citenamefont
  {Twitchen}, \citenamefont {Elkouss}, \citenamefont {Wehner}, \citenamefont
  {Taminiau},\ and\ \citenamefont {Hanson}}]{Hensen2015}%
  \BibitemOpen
  \bibfield  {author} {\bibinfo {author} {\bibfnamefont {B.}~\bibnamefont
  {Hensen}}, \bibinfo {author} {\bibfnamefont {H.}~\bibnamefont {Bernien}},
  \bibinfo {author} {\bibfnamefont {A.~E.}\ \bibnamefont {Dr{\'{e}}au}},
  \bibinfo {author} {\bibfnamefont {A.}~\bibnamefont {Reiserer}}, \bibinfo
  {author} {\bibfnamefont {N.}~\bibnamefont {Kalb}}, \bibinfo {author}
  {\bibfnamefont {M.~S.}\ \bibnamefont {Blok}}, \bibinfo {author}
  {\bibfnamefont {J.}~\bibnamefont {Ruitenberg}}, \bibinfo {author}
  {\bibfnamefont {R.~F.~L.}\ \bibnamefont {Vermeulen}}, \bibinfo {author}
  {\bibfnamefont {R.~N.}\ \bibnamefont {Schouten}}, \bibinfo {author}
  {\bibfnamefont {C.}~\bibnamefont {Abell{\'{a}}n}}, \bibinfo {author}
  {\bibfnamefont {W.}~\bibnamefont {Amaya}}, \bibinfo {author} {\bibfnamefont
  {V.}~\bibnamefont {Pruneri}}, \bibinfo {author} {\bibfnamefont {M.~W.}\
  \bibnamefont {Mitchell}}, \bibinfo {author} {\bibfnamefont {M.}~\bibnamefont
  {Markham}}, \bibinfo {author} {\bibfnamefont {D.~J.}\ \bibnamefont
  {Twitchen}}, \bibinfo {author} {\bibfnamefont {D.}~\bibnamefont {Elkouss}},
  \bibinfo {author} {\bibfnamefont {S.}~\bibnamefont {Wehner}}, \bibinfo
  {author} {\bibfnamefont {T.~H.}\ \bibnamefont {Taminiau}}, \ and\ \bibinfo
  {author} {\bibfnamefont {R.}~\bibnamefont {Hanson}},\ }\href {\doibase
  10.1038/nature15759} {\bibfield  {journal} {\bibinfo  {journal} {Nature}\
  }\textbf {\bibinfo {volume} {526}},\ \bibinfo {pages} {682} (\bibinfo {year}
  {2015})}\BibitemShut {NoStop}%
\bibitem [{\citenamefont {Lauritzen}\ \emph {et~al.}(2010)\citenamefont
  {Lauritzen}, \citenamefont {Min{\'{a}}ř}, \citenamefont {de~Riedmatten},
  \citenamefont {Afzelius}, \citenamefont {Sangouard}, \citenamefont {Simon},\
  and\ \citenamefont {Gisin}}]{Lauritzen2010}%
  \BibitemOpen
  \bibfield  {author} {\bibinfo {author} {\bibfnamefont {B.}~\bibnamefont
  {Lauritzen}}, \bibinfo {author} {\bibfnamefont {J.}~\bibnamefont
  {Min{\'{a}}ř}}, \bibinfo {author} {\bibfnamefont {H.}~\bibnamefont
  {de~Riedmatten}}, \bibinfo {author} {\bibfnamefont {M.}~\bibnamefont
  {Afzelius}}, \bibinfo {author} {\bibfnamefont {N.}~\bibnamefont {Sangouard}},
  \bibinfo {author} {\bibfnamefont {C.}~\bibnamefont {Simon}}, \ and\ \bibinfo
  {author} {\bibfnamefont {N.}~\bibnamefont {Gisin}},\ }\href {\doibase
  10.1103/PhysRevLett.104.080502} {\bibfield  {journal} {\bibinfo  {journal}
  {Physical Review Letters}\ }\textbf {\bibinfo {volume} {104}},\ \bibinfo
  {pages} {080502} (\bibinfo {year} {2010})}\BibitemShut {NoStop}%
\bibitem [{\citenamefont {Dajczgewand}\ \emph {et~al.}(2014)\citenamefont
  {Dajczgewand}, \citenamefont {{Le Gou{\"{e}}t}}, \citenamefont
  {Louchet-Chauvet},\ and\ \citenamefont {Chaneli{\`{e}}re}}]{Dajczgewand2014}%
  \BibitemOpen
  \bibfield  {author} {\bibinfo {author} {\bibfnamefont {J.}~\bibnamefont
  {Dajczgewand}}, \bibinfo {author} {\bibfnamefont {J.-L.}\ \bibnamefont {{Le
  Gou{\"{e}}t}}}, \bibinfo {author} {\bibfnamefont {A.}~\bibnamefont
  {Louchet-Chauvet}}, \ and\ \bibinfo {author} {\bibfnamefont {T.}~\bibnamefont
  {Chaneli{\`{e}}re}},\ }\href {\doibase 10.1364/OL.39.002711} {\bibfield
  {journal} {\bibinfo  {journal} {Optics Letters}\ }\textbf {\bibinfo {volume}
  {39}},\ \bibinfo {pages} {2711} (\bibinfo {year} {2014})}\BibitemShut
  {NoStop}%
\bibitem [{\citenamefont {Saglamyurek}\ \emph {et~al.}(2015)\citenamefont
  {Saglamyurek}, \citenamefont {Jin}, \citenamefont {Verma}, \citenamefont
  {Shaw}, \citenamefont {Marsili}, \citenamefont {Nam}, \citenamefont {Oblak},\
  and\ \citenamefont {Tittel}}]{Saglamyurek2015}%
  \BibitemOpen
  \bibfield  {author} {\bibinfo {author} {\bibfnamefont {E.}~\bibnamefont
  {Saglamyurek}}, \bibinfo {author} {\bibfnamefont {J.}~\bibnamefont {Jin}},
  \bibinfo {author} {\bibfnamefont {V.~B.}\ \bibnamefont {Verma}}, \bibinfo
  {author} {\bibfnamefont {M.~D.}\ \bibnamefont {Shaw}}, \bibinfo {author}
  {\bibfnamefont {F.}~\bibnamefont {Marsili}}, \bibinfo {author} {\bibfnamefont
  {S.~W.}\ \bibnamefont {Nam}}, \bibinfo {author} {\bibfnamefont
  {D.}~\bibnamefont {Oblak}}, \ and\ \bibinfo {author} {\bibfnamefont
  {W.}~\bibnamefont {Tittel}},\ }\href {\doibase 10.1038/nphoton.2014.311}
  {\bibfield  {journal} {\bibinfo  {journal} {Nature Photonics}\ }\textbf
  {\bibinfo {volume} {9}},\ \bibinfo {pages} {83} (\bibinfo {year}
  {2015})}\BibitemShut {NoStop}%
\bibitem [{\citenamefont {Simon}\ \emph {et~al.}(2007)\citenamefont {Simon},
  \citenamefont {de~Riedmatten}, \citenamefont {Afzelius}, \citenamefont
  {Sangouard}, \citenamefont {Zbinden},\ and\ \citenamefont
  {Gisin}}]{Simon2007}%
  \BibitemOpen
  \bibfield  {author} {\bibinfo {author} {\bibfnamefont {C.}~\bibnamefont
  {Simon}}, \bibinfo {author} {\bibfnamefont {H.}~\bibnamefont
  {de~Riedmatten}}, \bibinfo {author} {\bibfnamefont {M.}~\bibnamefont
  {Afzelius}}, \bibinfo {author} {\bibfnamefont {N.}~\bibnamefont {Sangouard}},
  \bibinfo {author} {\bibfnamefont {H.}~\bibnamefont {Zbinden}}, \ and\
  \bibinfo {author} {\bibfnamefont {N.}~\bibnamefont {Gisin}},\ }\href
  {\doibase 10.1103/PhysRevLett.98.190503} {\bibfield  {journal} {\bibinfo
  {journal} {Physical Review Letters}\ }\textbf {\bibinfo {volume} {98}},\
  \bibinfo {pages} {190503} (\bibinfo {year} {2007})}\BibitemShut {NoStop}%
\bibitem [{\citenamefont {Clausen}\ \emph {et~al.}(2011)\citenamefont
  {Clausen}, \citenamefont {Usmani}, \citenamefont {Bussi{\`{e}}res},
  \citenamefont {Sangouard}, \citenamefont {Afzelius}, \citenamefont
  {de~Riedmatten},\ and\ \citenamefont {Gisin}}]{Clausen2011}%
  \BibitemOpen
  \bibfield  {author} {\bibinfo {author} {\bibfnamefont {C.}~\bibnamefont
  {Clausen}}, \bibinfo {author} {\bibfnamefont {I.}~\bibnamefont {Usmani}},
  \bibinfo {author} {\bibfnamefont {F.}~\bibnamefont {Bussi{\`{e}}res}},
  \bibinfo {author} {\bibfnamefont {N.}~\bibnamefont {Sangouard}}, \bibinfo
  {author} {\bibfnamefont {M.}~\bibnamefont {Afzelius}}, \bibinfo {author}
  {\bibfnamefont {H.}~\bibnamefont {de~Riedmatten}}, \ and\ \bibinfo {author}
  {\bibfnamefont {N.}~\bibnamefont {Gisin}},\ }\href {\doibase
  10.1038/nature09662} {\bibfield  {journal} {\bibinfo  {journal} {Nature}\
  }\textbf {\bibinfo {volume} {469}},\ \bibinfo {pages} {508} (\bibinfo {year}
  {2011})}\BibitemShut {NoStop}%
\bibitem [{\citenamefont {Saglamyurek}\ \emph {et~al.}(2011)\citenamefont
  {Saglamyurek}, \citenamefont {Sinclair}, \citenamefont {Jin}, \citenamefont
  {Slater}, \citenamefont {Oblak}, \citenamefont {Bussi{\`{e}}res},
  \citenamefont {George}, \citenamefont {Ricken}, \citenamefont {Sohler},\ and\
  \citenamefont {Tittel}}]{Saglamyurek2011}%
  \BibitemOpen
  \bibfield  {author} {\bibinfo {author} {\bibfnamefont {E.}~\bibnamefont
  {Saglamyurek}}, \bibinfo {author} {\bibfnamefont {N.}~\bibnamefont
  {Sinclair}}, \bibinfo {author} {\bibfnamefont {J.}~\bibnamefont {Jin}},
  \bibinfo {author} {\bibfnamefont {J.~A.}\ \bibnamefont {Slater}}, \bibinfo
  {author} {\bibfnamefont {D.}~\bibnamefont {Oblak}}, \bibinfo {author}
  {\bibfnamefont {F.}~\bibnamefont {Bussi{\`{e}}res}}, \bibinfo {author}
  {\bibfnamefont {M.}~\bibnamefont {George}}, \bibinfo {author} {\bibfnamefont
  {R.}~\bibnamefont {Ricken}}, \bibinfo {author} {\bibfnamefont
  {W.}~\bibnamefont {Sohler}}, \ and\ \bibinfo {author} {\bibfnamefont
  {W.}~\bibnamefont {Tittel}},\ }\href {\doibase 10.1038/nature09719}
  {\bibfield  {journal} {\bibinfo  {journal} {Nature}\ }\textbf {\bibinfo
  {volume} {469}},\ \bibinfo {pages} {512} (\bibinfo {year}
  {2011})}\BibitemShut {NoStop}%
\bibitem [{\citenamefont {Riel{\"{a}}nder}\ \emph {et~al.}(2014)\citenamefont
  {Riel{\"{a}}nder}, \citenamefont {Kutluer}, \citenamefont {Ledingham},
  \citenamefont {G{\"{u}}ndoğan}, \citenamefont {Fekete}, \citenamefont
  {Mazzera},\ and\ \citenamefont {de~Riedmatten}}]{Rielander2014}%
  \BibitemOpen
  \bibfield  {author} {\bibinfo {author} {\bibfnamefont {D.}~\bibnamefont
  {Riel{\"{a}}nder}}, \bibinfo {author} {\bibfnamefont {K.}~\bibnamefont
  {Kutluer}}, \bibinfo {author} {\bibfnamefont {P.~M.}\ \bibnamefont
  {Ledingham}}, \bibinfo {author} {\bibfnamefont {M.}~\bibnamefont
  {G{\"{u}}ndoğan}}, \bibinfo {author} {\bibfnamefont {J.}~\bibnamefont
  {Fekete}}, \bibinfo {author} {\bibfnamefont {M.}~\bibnamefont {Mazzera}}, \
  and\ \bibinfo {author} {\bibfnamefont {H.}~\bibnamefont {de~Riedmatten}},\
  }\href {\doibase 10.1103/PhysRevLett.112.040504} {\bibfield  {journal}
  {\bibinfo  {journal} {Physical Review Letters}\ }\textbf {\bibinfo {volume}
  {112}},\ \bibinfo {pages} {040504} (\bibinfo {year} {2014})}\BibitemShut
  {NoStop}%
\bibitem [{\citenamefont {Bussi{\`{e}}res}\ \emph {et~al.}(2014)\citenamefont
  {Bussi{\`{e}}res}, \citenamefont {Clausen}, \citenamefont {Tiranov},
  \citenamefont {Korzh}, \citenamefont {Verma}, \citenamefont {Nam},
  \citenamefont {Marsili}, \citenamefont {Ferrier}, \citenamefont {Goldner},
  \citenamefont {Herrmann}, \citenamefont {Silberhorn}, \citenamefont {Sohler},
  \citenamefont {Afzelius},\ and\ \citenamefont {Gisin}}]{Bussieres2014}%
  \BibitemOpen
  \bibfield  {author} {\bibinfo {author} {\bibfnamefont {F.}~\bibnamefont
  {Bussi{\`{e}}res}}, \bibinfo {author} {\bibfnamefont {C.}~\bibnamefont
  {Clausen}}, \bibinfo {author} {\bibfnamefont {A.}~\bibnamefont {Tiranov}},
  \bibinfo {author} {\bibfnamefont {B.}~\bibnamefont {Korzh}}, \bibinfo
  {author} {\bibfnamefont {V.~B.}\ \bibnamefont {Verma}}, \bibinfo {author}
  {\bibfnamefont {S.~W.}\ \bibnamefont {Nam}}, \bibinfo {author} {\bibfnamefont
  {F.}~\bibnamefont {Marsili}}, \bibinfo {author} {\bibfnamefont
  {A.}~\bibnamefont {Ferrier}}, \bibinfo {author} {\bibfnamefont
  {P.}~\bibnamefont {Goldner}}, \bibinfo {author} {\bibfnamefont
  {H.}~\bibnamefont {Herrmann}}, \bibinfo {author} {\bibfnamefont
  {C.}~\bibnamefont {Silberhorn}}, \bibinfo {author} {\bibfnamefont
  {W.}~\bibnamefont {Sohler}}, \bibinfo {author} {\bibfnamefont
  {M.}~\bibnamefont {Afzelius}}, \ and\ \bibinfo {author} {\bibfnamefont
  {N.}~\bibnamefont {Gisin}},\ }\href {\doibase 10.1038/nphoton.2014.215}
  {\bibfield  {journal} {\bibinfo  {journal} {Nature Photonics}\ }\textbf
  {\bibinfo {volume} {8}},\ \bibinfo {pages} {775} (\bibinfo {year}
  {2014})}\BibitemShut {NoStop}%
\bibitem [{\citenamefont {Lenhard}\ \emph {et~al.}(2015)\citenamefont
  {Lenhard}, \citenamefont {Bock}, \citenamefont {Becher}, \citenamefont
  {Kucera}, \citenamefont {Brito}, \citenamefont {Eich}, \citenamefont
  {M{\"{u}}ller},\ and\ \citenamefont {Eschner}}]{Lenhard2015}%
  \BibitemOpen
  \bibfield  {author} {\bibinfo {author} {\bibfnamefont {A.}~\bibnamefont
  {Lenhard}}, \bibinfo {author} {\bibfnamefont {M.}~\bibnamefont {Bock}},
  \bibinfo {author} {\bibfnamefont {C.}~\bibnamefont {Becher}}, \bibinfo
  {author} {\bibfnamefont {S.}~\bibnamefont {Kucera}}, \bibinfo {author}
  {\bibfnamefont {J.}~\bibnamefont {Brito}}, \bibinfo {author} {\bibfnamefont
  {P.}~\bibnamefont {Eich}}, \bibinfo {author} {\bibfnamefont {P.}~\bibnamefont
  {M{\"{u}}ller}}, \ and\ \bibinfo {author} {\bibfnamefont {J.}~\bibnamefont
  {Eschner}},\ }\href {\doibase 10.1103/PhysRevA.92.063827} {\bibfield
  {journal} {\bibinfo  {journal} {Physical Review A}\ }\textbf {\bibinfo
  {volume} {92}},\ \bibinfo {pages} {063827} (\bibinfo {year}
  {2015})}\BibitemShut {NoStop}%
\bibitem [{\citenamefont {Schunk}\ \emph {et~al.}(2015)\citenamefont {Schunk},
  \citenamefont {Vogl}, \citenamefont {Strekalov}, \citenamefont
  {F{\"{o}}rtsch}, \citenamefont {Sedlmeir}, \citenamefont {Schwefel},
  \citenamefont {G{\"{o}}belt}, \citenamefont {Christiansen}, \citenamefont
  {Leuchs},\ and\ \citenamefont {Marquardt}}]{Schunk2015}%
  \BibitemOpen
  \bibfield  {author} {\bibinfo {author} {\bibfnamefont {G.}~\bibnamefont
  {Schunk}}, \bibinfo {author} {\bibfnamefont {U.}~\bibnamefont {Vogl}},
  \bibinfo {author} {\bibfnamefont {D.~V.}\ \bibnamefont {Strekalov}}, \bibinfo
  {author} {\bibfnamefont {M.}~\bibnamefont {F{\"{o}}rtsch}}, \bibinfo {author}
  {\bibfnamefont {F.}~\bibnamefont {Sedlmeir}}, \bibinfo {author}
  {\bibfnamefont {H.~G.~L.}\ \bibnamefont {Schwefel}}, \bibinfo {author}
  {\bibfnamefont {M.}~\bibnamefont {G{\"{o}}belt}}, \bibinfo {author}
  {\bibfnamefont {S.}~\bibnamefont {Christiansen}}, \bibinfo {author}
  {\bibfnamefont {G.}~\bibnamefont {Leuchs}}, \ and\ \bibinfo {author}
  {\bibfnamefont {C.}~\bibnamefont {Marquardt}},\ }\href {\doibase
  10.1364/OPTICA.2.000773} {\bibfield  {journal} {\bibinfo  {journal} {Optica}\
  }\textbf {\bibinfo {volume} {2}},\ \bibinfo {pages} {773} (\bibinfo {year}
  {2015})}\BibitemShut {NoStop}%
\bibitem [{\citenamefont {Zhang}\ \emph {et~al.}(2016)\citenamefont {Zhang},
  \citenamefont {Ding}, \citenamefont {Shi}, \citenamefont {Li}, \citenamefont
  {Zhou}, \citenamefont {Shi},\ and\ \citenamefont {Guo}}]{Zhang2016}%
  \BibitemOpen
  \bibfield  {author} {\bibinfo {author} {\bibfnamefont {W.}~\bibnamefont
  {Zhang}}, \bibinfo {author} {\bibfnamefont {D.-S.}\ \bibnamefont {Ding}},
  \bibinfo {author} {\bibfnamefont {S.}~\bibnamefont {Shi}}, \bibinfo {author}
  {\bibfnamefont {Y.}~\bibnamefont {Li}}, \bibinfo {author} {\bibfnamefont
  {Z.-Y.}\ \bibnamefont {Zhou}}, \bibinfo {author} {\bibfnamefont {B.-S.}\
  \bibnamefont {Shi}}, \ and\ \bibinfo {author} {\bibfnamefont {G.-C.}\
  \bibnamefont {Guo}},\ }\href {\doibase 10.1103/PhysRevA.93.022316} {\bibfield
   {journal} {\bibinfo  {journal} {Physical Review A}\ }\textbf {\bibinfo
  {volume} {93}},\ \bibinfo {pages} {022316} (\bibinfo {year}
  {2016})}\BibitemShut {NoStop}%
\bibitem [{\citenamefont {Tanzilli}\ \emph {et~al.}(2005)\citenamefont
  {Tanzilli}, \citenamefont {Tittel}, \citenamefont {Halder}, \citenamefont
  {Alibart}, \citenamefont {Baldi}, \citenamefont {Gisin},\ and\ \citenamefont
  {Zbinden}}]{Tanzilli2005}%
  \BibitemOpen
  \bibfield  {author} {\bibinfo {author} {\bibfnamefont {S.}~\bibnamefont
  {Tanzilli}}, \bibinfo {author} {\bibfnamefont {W.}~\bibnamefont {Tittel}},
  \bibinfo {author} {\bibfnamefont {M.}~\bibnamefont {Halder}}, \bibinfo
  {author} {\bibfnamefont {O.}~\bibnamefont {Alibart}}, \bibinfo {author}
  {\bibfnamefont {P.}~\bibnamefont {Baldi}}, \bibinfo {author} {\bibfnamefont
  {N.}~\bibnamefont {Gisin}}, \ and\ \bibinfo {author} {\bibfnamefont
  {H.}~\bibnamefont {Zbinden}},\ }\href {\doibase 10.1038/nature04009}
  {\bibfield  {journal} {\bibinfo  {journal} {Nature}\ }\textbf {\bibinfo
  {volume} {437}},\ \bibinfo {pages} {116} (\bibinfo {year}
  {2005})}\BibitemShut {NoStop}%
\bibitem [{\citenamefont {Radnaev}\ \emph {et~al.}(2010)\citenamefont
  {Radnaev}, \citenamefont {Dudin}, \citenamefont {Zhao}, \citenamefont {Jen},
  \citenamefont {Jenkins}, \citenamefont {Kuzmich},\ and\ \citenamefont
  {Kennedy}}]{Radnaev2010}%
  \BibitemOpen
  \bibfield  {author} {\bibinfo {author} {\bibfnamefont {A.~G.}\ \bibnamefont
  {Radnaev}}, \bibinfo {author} {\bibfnamefont {Y.~O.}\ \bibnamefont {Dudin}},
  \bibinfo {author} {\bibfnamefont {R.}~\bibnamefont {Zhao}}, \bibinfo {author}
  {\bibfnamefont {H.~H.}\ \bibnamefont {Jen}}, \bibinfo {author} {\bibfnamefont
  {S.~D.}\ \bibnamefont {Jenkins}}, \bibinfo {author} {\bibfnamefont
  {A.}~\bibnamefont {Kuzmich}}, \ and\ \bibinfo {author} {\bibfnamefont
  {T.~A.~B.}\ \bibnamefont {Kennedy}},\ }\href {\doibase 10.1038/nphys1773}
  {\bibfield  {journal} {\bibinfo  {journal} {Nature Physics}\ }\textbf
  {\bibinfo {volume} {6}},\ \bibinfo {pages} {894} (\bibinfo {year}
  {2010})}\BibitemShut {NoStop}%
\bibitem [{\citenamefont {Dudin}\ \emph {et~al.}(2010)\citenamefont {Dudin},
  \citenamefont {Radnaev}, \citenamefont {Zhao}, \citenamefont {Blumoff},
  \citenamefont {Kennedy},\ and\ \citenamefont {Kuzmich}}]{Dudin2010}%
  \BibitemOpen
  \bibfield  {author} {\bibinfo {author} {\bibfnamefont {Y.~O.}\ \bibnamefont
  {Dudin}}, \bibinfo {author} {\bibfnamefont {A.~G.}\ \bibnamefont {Radnaev}},
  \bibinfo {author} {\bibfnamefont {R.}~\bibnamefont {Zhao}}, \bibinfo {author}
  {\bibfnamefont {J.~Z.}\ \bibnamefont {Blumoff}}, \bibinfo {author}
  {\bibfnamefont {T.~A.~B.}\ \bibnamefont {Kennedy}}, \ and\ \bibinfo {author}
  {\bibfnamefont {A.}~\bibnamefont {Kuzmich}},\ }\href {\doibase
  10.1103/PhysRevLett.105.260502} {\bibfield  {journal} {\bibinfo  {journal}
  {Physical Review Letters}\ }\textbf {\bibinfo {volume} {105}},\ \bibinfo
  {pages} {260502} (\bibinfo {year} {2010})}\BibitemShut {NoStop}%
\bibitem [{\citenamefont {Ikuta}\ \emph {et~al.}(2011)\citenamefont {Ikuta},
  \citenamefont {Kusaka}, \citenamefont {Kitano}, \citenamefont {Kato},
  \citenamefont {Yamamoto}, \citenamefont {Koashi},\ and\ \citenamefont
  {Imoto}}]{Ikuta2011}%
  \BibitemOpen
  \bibfield  {author} {\bibinfo {author} {\bibfnamefont {R.}~\bibnamefont
  {Ikuta}}, \bibinfo {author} {\bibfnamefont {Y.}~\bibnamefont {Kusaka}},
  \bibinfo {author} {\bibfnamefont {T.}~\bibnamefont {Kitano}}, \bibinfo
  {author} {\bibfnamefont {H.}~\bibnamefont {Kato}}, \bibinfo {author}
  {\bibfnamefont {T.}~\bibnamefont {Yamamoto}}, \bibinfo {author}
  {\bibfnamefont {M.}~\bibnamefont {Koashi}}, \ and\ \bibinfo {author}
  {\bibfnamefont {N.}~\bibnamefont {Imoto}},\ }\href {\doibase
  10.1038/ncomms1544} {\bibfield  {journal} {\bibinfo  {journal} {Nature
  Communications}\ }\textbf {\bibinfo {volume} {2}},\ \bibinfo {pages} {1544}
  (\bibinfo {year} {2011})}\BibitemShut {NoStop}%
\bibitem [{\citenamefont {Zaske}\ \emph {et~al.}(2012)\citenamefont {Zaske},
  \citenamefont {Lenhard}, \citenamefont {Ke{\ss}ler}, \citenamefont {Kettler},
  \citenamefont {Hepp}, \citenamefont {Arend}, \citenamefont {Albrecht},
  \citenamefont {Schulz}, \citenamefont {Jetter}, \citenamefont {Michler},\
  and\ \citenamefont {Becher}}]{Zaske2012}%
  \BibitemOpen
  \bibfield  {author} {\bibinfo {author} {\bibfnamefont {S.}~\bibnamefont
  {Zaske}}, \bibinfo {author} {\bibfnamefont {A.}~\bibnamefont {Lenhard}},
  \bibinfo {author} {\bibfnamefont {C.~A.}\ \bibnamefont {Ke{\ss}ler}},
  \bibinfo {author} {\bibfnamefont {J.}~\bibnamefont {Kettler}}, \bibinfo
  {author} {\bibfnamefont {C.}~\bibnamefont {Hepp}}, \bibinfo {author}
  {\bibfnamefont {C.}~\bibnamefont {Arend}}, \bibinfo {author} {\bibfnamefont
  {R.}~\bibnamefont {Albrecht}}, \bibinfo {author} {\bibfnamefont {W.-M.}\
  \bibnamefont {Schulz}}, \bibinfo {author} {\bibfnamefont {M.}~\bibnamefont
  {Jetter}}, \bibinfo {author} {\bibfnamefont {P.}~\bibnamefont {Michler}}, \
  and\ \bibinfo {author} {\bibfnamefont {C.}~\bibnamefont {Becher}},\ }\href
  {\doibase 10.1103/PhysRevLett.109.147404} {\bibfield  {journal} {\bibinfo
  {journal} {Physical Review Letters}\ }\textbf {\bibinfo {volume} {109}},\
  \bibinfo {pages} {147404} (\bibinfo {year} {2012})}\BibitemShut {NoStop}%
\bibitem [{\citenamefont {Ates}\ \emph {et~al.}(2012)\citenamefont {Ates},
  \citenamefont {Agha}, \citenamefont {Gulinatti}, \citenamefont {Rech},
  \citenamefont {Rakher}, \citenamefont {Badolato},\ and\ \citenamefont
  {Srinivasan}}]{Ates2012}%
  \BibitemOpen
  \bibfield  {author} {\bibinfo {author} {\bibfnamefont {S.}~\bibnamefont
  {Ates}}, \bibinfo {author} {\bibfnamefont {I.}~\bibnamefont {Agha}}, \bibinfo
  {author} {\bibfnamefont {A.}~\bibnamefont {Gulinatti}}, \bibinfo {author}
  {\bibfnamefont {I.}~\bibnamefont {Rech}}, \bibinfo {author} {\bibfnamefont
  {M.~T.}\ \bibnamefont {Rakher}}, \bibinfo {author} {\bibfnamefont
  {A.}~\bibnamefont {Badolato}}, \ and\ \bibinfo {author} {\bibfnamefont
  {K.}~\bibnamefont {Srinivasan}},\ }\href {\doibase
  10.1103/PhysRevLett.109.147405} {\bibfield  {journal} {\bibinfo  {journal}
  {Physical review letters}\ }\textbf {\bibinfo {volume} {109}},\ \bibinfo
  {pages} {147405} (\bibinfo {year} {2012})}\BibitemShut {NoStop}%
\bibitem [{\citenamefont {Ikuta}\ \emph {et~al.}(2013)\citenamefont {Ikuta},
  \citenamefont {Kato}, \citenamefont {Kusaka}, \citenamefont {Miki},
  \citenamefont {Yamashita}, \citenamefont {Terai}, \citenamefont {Fujiwara},
  \citenamefont {Yamamoto}, \citenamefont {Koashi}, \citenamefont {Sasaki},
  \citenamefont {Wang},\ and\ \citenamefont {Imoto}}]{Ikuta2013}%
  \BibitemOpen
  \bibfield  {author} {\bibinfo {author} {\bibfnamefont {R.}~\bibnamefont
  {Ikuta}}, \bibinfo {author} {\bibfnamefont {H.}~\bibnamefont {Kato}},
  \bibinfo {author} {\bibfnamefont {Y.}~\bibnamefont {Kusaka}}, \bibinfo
  {author} {\bibfnamefont {S.}~\bibnamefont {Miki}}, \bibinfo {author}
  {\bibfnamefont {T.}~\bibnamefont {Yamashita}}, \bibinfo {author}
  {\bibfnamefont {H.}~\bibnamefont {Terai}}, \bibinfo {author} {\bibfnamefont
  {M.}~\bibnamefont {Fujiwara}}, \bibinfo {author} {\bibfnamefont
  {T.}~\bibnamefont {Yamamoto}}, \bibinfo {author} {\bibfnamefont
  {M.}~\bibnamefont {Koashi}}, \bibinfo {author} {\bibfnamefont
  {M.}~\bibnamefont {Sasaki}}, \bibinfo {author} {\bibfnamefont
  {Z.}~\bibnamefont {Wang}}, \ and\ \bibinfo {author} {\bibfnamefont
  {N.}~\bibnamefont {Imoto}},\ }\href {\doibase 10.1103/PhysRevA.87.010301}
  {\bibfield  {journal} {\bibinfo  {journal} {Physical Review A}\ }\textbf
  {\bibinfo {volume} {87}},\ \bibinfo {pages} {010301} (\bibinfo {year}
  {2013})}\BibitemShut {NoStop}%
\bibitem [{\citenamefont {Albrecht}\ \emph {et~al.}(2014)\citenamefont
  {Albrecht}, \citenamefont {Farrera}, \citenamefont {Fernandez-Gonzalvo},
  \citenamefont {Cristiani},\ and\ \citenamefont
  {de~Riedmatten}}]{Albrecht2014}%
  \BibitemOpen
  \bibfield  {author} {\bibinfo {author} {\bibfnamefont {B.}~\bibnamefont
  {Albrecht}}, \bibinfo {author} {\bibfnamefont {P.}~\bibnamefont {Farrera}},
  \bibinfo {author} {\bibfnamefont {X.}~\bibnamefont {Fernandez-Gonzalvo}},
  \bibinfo {author} {\bibfnamefont {M.}~\bibnamefont {Cristiani}}, \ and\
  \bibinfo {author} {\bibfnamefont {H.}~\bibnamefont {de~Riedmatten}},\ }\href
  {\doibase 10.1038/ncomms4376} {\bibfield  {journal} {\bibinfo  {journal}
  {Nature Communications}\ }\textbf {\bibinfo {volume} {5}},\ \bibinfo {pages}
  {3376} (\bibinfo {year} {2014})}\BibitemShut {NoStop}%
\bibitem [{\citenamefont {Maring}\ \emph {et~al.}(2014)\citenamefont {Maring},
  \citenamefont {Kutluer}, \citenamefont {Cohen}, \citenamefont {Cristiani},
  \citenamefont {Mazzera}, \citenamefont {Ledingham},\ and\ \citenamefont
  {de~Riedmatten}}]{Maring2014}%
  \BibitemOpen
  \bibfield  {author} {\bibinfo {author} {\bibfnamefont {N.}~\bibnamefont
  {Maring}}, \bibinfo {author} {\bibfnamefont {K.}~\bibnamefont {Kutluer}},
  \bibinfo {author} {\bibfnamefont {J.}~\bibnamefont {Cohen}}, \bibinfo
  {author} {\bibfnamefont {M.}~\bibnamefont {Cristiani}}, \bibinfo {author}
  {\bibfnamefont {M.}~\bibnamefont {Mazzera}}, \bibinfo {author} {\bibfnamefont
  {P.~M.}\ \bibnamefont {Ledingham}}, \ and\ \bibinfo {author} {\bibfnamefont
  {H.}~\bibnamefont {de~Riedmatten}},\ }\href {\doibase
  10.1088/1367-2630/16/11/113021} {\bibfield  {journal} {\bibinfo  {journal}
  {New Journal of Physics}\ }\textbf {\bibinfo {volume} {16}},\ \bibinfo
  {pages} {113021} (\bibinfo {year} {2014})}\BibitemShut {NoStop}%
\bibitem [{\citenamefont {Duan}\ \emph {et~al.}(2001)\citenamefont {Duan},
  \citenamefont {Lukin}, \citenamefont {Cirac},\ and\ \citenamefont
  {Zoller}}]{Duan2001}%
  \BibitemOpen
  \bibfield  {author} {\bibinfo {author} {\bibfnamefont {L.~M.}\ \bibnamefont
  {Duan}}, \bibinfo {author} {\bibfnamefont {M.~D.}\ \bibnamefont {Lukin}},
  \bibinfo {author} {\bibfnamefont {J.~I.}\ \bibnamefont {Cirac}}, \ and\
  \bibinfo {author} {\bibfnamefont {P.}~\bibnamefont {Zoller}},\ }\href
  {\doibase 10.1038/35106500} {\bibfield  {journal} {\bibinfo  {journal}
  {Nature}\ }\textbf {\bibinfo {volume} {414}},\ \bibinfo {pages} {413}
  (\bibinfo {year} {2001})}\BibitemShut {NoStop}%
\bibitem [{\citenamefont {Fernandez-Gonzalvo}\ \emph
  {et~al.}(2013)\citenamefont {Fernandez-Gonzalvo}, \citenamefont {Corrielli},
  \citenamefont {Albrecht}, \citenamefont {Grimau}, \citenamefont {Cristiani},\
  and\ \citenamefont {de~Riedmatten}}]{Fernandez-Gonzalvo2013}%
  \BibitemOpen
  \bibfield  {author} {\bibinfo {author} {\bibfnamefont {X.}~\bibnamefont
  {Fernandez-Gonzalvo}}, \bibinfo {author} {\bibfnamefont {G.}~\bibnamefont
  {Corrielli}}, \bibinfo {author} {\bibfnamefont {B.}~\bibnamefont {Albrecht}},
  \bibinfo {author} {\bibfnamefont {M.}~\bibnamefont {Grimau}}, \bibinfo
  {author} {\bibfnamefont {M.}~\bibnamefont {Cristiani}}, \ and\ \bibinfo
  {author} {\bibfnamefont {H.}~\bibnamefont {de~Riedmatten}},\ }\href {\doibase
  10.1364/OE.21.019473} {\bibfield  {journal} {\bibinfo  {journal} {Optics
  Express}\ }\textbf {\bibinfo {volume} {21}},\ \bibinfo {pages} {19473}
  (\bibinfo {year} {2013})}\BibitemShut {NoStop}%
\bibitem [{\citenamefont {Sekatski}\ \emph {et~al.}(2012)\citenamefont
  {Sekatski}, \citenamefont {Sangouard}, \citenamefont {Bussi{\`{e}}res},
  \citenamefont {Clausen}, \citenamefont {Gisin},\ and\ \citenamefont
  {Zbinden}}]{Sekatski2012}%
  \BibitemOpen
  \bibfield  {author} {\bibinfo {author} {\bibfnamefont {P.}~\bibnamefont
  {Sekatski}}, \bibinfo {author} {\bibfnamefont {N.}~\bibnamefont {Sangouard}},
  \bibinfo {author} {\bibfnamefont {F.}~\bibnamefont {Bussi{\`{e}}res}},
  \bibinfo {author} {\bibfnamefont {C.}~\bibnamefont {Clausen}}, \bibinfo
  {author} {\bibfnamefont {N.}~\bibnamefont {Gisin}}, \ and\ \bibinfo {author}
  {\bibfnamefont {H.}~\bibnamefont {Zbinden}},\ }\href {\doibase
  10.1088/0953-4075/45/12/124016} {\bibfield  {journal} {\bibinfo  {journal}
  {Journal of Physics B: Atomic, Molecular and Optical Physics}\ }\textbf
  {\bibinfo {volume} {45}},\ \bibinfo {pages} {124016} (\bibinfo {year}
  {2012})}\BibitemShut {NoStop}%
\bibitem [{\citenamefont {de~Riedmatten}\ \emph {et~al.}(2006)\citenamefont
  {de~Riedmatten}, \citenamefont {Laurat}, \citenamefont {Chou}, \citenamefont
  {Schomburg}, \citenamefont {Felinto},\ and\ \citenamefont
  {Kimble}}]{DeRiedmatten2006}%
  \BibitemOpen
  \bibfield  {author} {\bibinfo {author} {\bibfnamefont {H.}~\bibnamefont
  {de~Riedmatten}}, \bibinfo {author} {\bibfnamefont {J.}~\bibnamefont
  {Laurat}}, \bibinfo {author} {\bibfnamefont {C.~W.}\ \bibnamefont {Chou}},
  \bibinfo {author} {\bibfnamefont {E.~W.}\ \bibnamefont {Schomburg}}, \bibinfo
  {author} {\bibfnamefont {D.}~\bibnamefont {Felinto}}, \ and\ \bibinfo
  {author} {\bibfnamefont {H.~J.}\ \bibnamefont {Kimble}},\ }\href {\doibase
  10.1103/PhysRevLett.97.113603} {\bibfield  {journal} {\bibinfo  {journal}
  {Physical Review Letters}\ }\textbf {\bibinfo {volume} {97}},\ \bibinfo
  {pages} {113603} (\bibinfo {year} {2006})}\BibitemShut {NoStop}%
\bibitem [{\citenamefont {Albrecht}\ \emph {et~al.}(2015)\citenamefont
  {Albrecht}, \citenamefont {Farrera}, \citenamefont {Heinze}, \citenamefont
  {Cristiani},\ and\ \citenamefont {de~Riedmatten}}]{Albrecht2015}%
  \BibitemOpen
  \bibfield  {author} {\bibinfo {author} {\bibfnamefont {B.}~\bibnamefont
  {Albrecht}}, \bibinfo {author} {\bibfnamefont {P.}~\bibnamefont {Farrera}},
  \bibinfo {author} {\bibfnamefont {G.}~\bibnamefont {Heinze}}, \bibinfo
  {author} {\bibfnamefont {M.}~\bibnamefont {Cristiani}}, \ and\ \bibinfo
  {author} {\bibfnamefont {H.}~\bibnamefont {de~Riedmatten}},\ }\href {\doibase
  10.1103/PhysRevLett.115.160501} {\bibfield  {journal} {\bibinfo  {journal}
  {Physical Review Letters}\ }\textbf {\bibinfo {volume} {115}},\ \bibinfo
  {pages} {160501} (\bibinfo {year} {2015})}\BibitemShut {NoStop}%
\bibitem [{\citenamefont {Laurat}\ \emph {et~al.}(2007)\citenamefont {Laurat},
  \citenamefont {Choi}, \citenamefont {Deng}, \citenamefont {Chou},\ and\
  \citenamefont {Kimble}}]{Laurat2007}%
  \BibitemOpen
  \bibfield  {author} {\bibinfo {author} {\bibfnamefont {J.}~\bibnamefont
  {Laurat}}, \bibinfo {author} {\bibfnamefont {K.~S.}\ \bibnamefont {Choi}},
  \bibinfo {author} {\bibfnamefont {H.}~\bibnamefont {Deng}}, \bibinfo {author}
  {\bibfnamefont {C.~W.}\ \bibnamefont {Chou}}, \ and\ \bibinfo {author}
  {\bibfnamefont {H.~J.}\ \bibnamefont {Kimble}},\ }\href {\doibase
  10.1103/PhysRevLett.99.180504} {\bibfield  {journal} {\bibinfo  {journal}
  {Physical Review Letters}\ }\textbf {\bibinfo {volume} {99}},\ \bibinfo
  {pages} {180504} (\bibinfo {year} {2007})}\BibitemShut {NoStop}%
\bibitem [{\citenamefont {Bao}\ \emph {et~al.}(2012)\citenamefont {Bao},
  \citenamefont {Reingruber}, \citenamefont {Dietrich}, \citenamefont {Rui},
  \citenamefont {D{\"{u}}ck}, \citenamefont {Strassel}, \citenamefont {Li},
  \citenamefont {Liu}, \citenamefont {Zhao},\ and\ \citenamefont
  {Pan}}]{Bao2012}%
  \BibitemOpen
  \bibfield  {author} {\bibinfo {author} {\bibfnamefont {X.-H.}\ \bibnamefont
  {Bao}}, \bibinfo {author} {\bibfnamefont {A.}~\bibnamefont {Reingruber}},
  \bibinfo {author} {\bibfnamefont {P.}~\bibnamefont {Dietrich}}, \bibinfo
  {author} {\bibfnamefont {J.}~\bibnamefont {Rui}}, \bibinfo {author}
  {\bibfnamefont {A.}~\bibnamefont {D{\"{u}}ck}}, \bibinfo {author}
  {\bibfnamefont {T.}~\bibnamefont {Strassel}}, \bibinfo {author}
  {\bibfnamefont {L.}~\bibnamefont {Li}}, \bibinfo {author} {\bibfnamefont
  {N.-L.}\ \bibnamefont {Liu}}, \bibinfo {author} {\bibfnamefont
  {B.}~\bibnamefont {Zhao}}, \ and\ \bibinfo {author} {\bibfnamefont {J.-W.}\
  \bibnamefont {Pan}},\ }\href {\doibase 10.1038/nphys2324} {\bibfield
  {journal} {\bibinfo  {journal} {Nature Physics}\ }\textbf {\bibinfo {volume}
  {8}},\ \bibinfo {pages} {517} (\bibinfo {year} {2012})}\BibitemShut {NoStop}%
\bibitem [{\citenamefont {Yang}\ \emph {et~al.}(2015)\citenamefont {Yang},
  \citenamefont {Wang}, \citenamefont {Bao},\ and\ \citenamefont
  {Pan}}]{Yang2016}%
  \BibitemOpen
  \bibfield  {author} {\bibinfo {author} {\bibfnamefont {S.-J.}\ \bibnamefont
  {Yang}}, \bibinfo {author} {\bibfnamefont {X.-J.}\ \bibnamefont {Wang}},
  \bibinfo {author} {\bibfnamefont {X.-H.}\ \bibnamefont {Bao}}, \ and\
  \bibinfo {author} {\bibfnamefont {J.-W.}\ \bibnamefont {Pan}},\ }\href
  {\doibase 10.1038/nphoton.2016.51} {\bibfield  {journal} {\bibinfo  {journal}
  {Nature Photonics}\ }\textbf {\bibinfo {volume} {10}},\ \bibinfo {pages}
  {381} (\bibinfo {year} {2015})}\BibitemShut {NoStop}%
\bibitem [{\citenamefont {Jiang}\ \emph {et~al.}(2016)\citenamefont {Jiang},
  \citenamefont {Rui}, \citenamefont {Bao},\ and\ \citenamefont
  {Pan}}]{Jiang2016}%
  \BibitemOpen
  \bibfield  {author} {\bibinfo {author} {\bibfnamefont {Y.}~\bibnamefont
  {Jiang}}, \bibinfo {author} {\bibfnamefont {J.}~\bibnamefont {Rui}}, \bibinfo
  {author} {\bibfnamefont {X.-H.}\ \bibnamefont {Bao}}, \ and\ \bibinfo
  {author} {\bibfnamefont {J.-W.}\ \bibnamefont {Pan}},\ }\href {\doibase
  10.1103/PhysRevA.93.063819} {\bibfield  {journal} {\bibinfo  {journal}
  {Physical Review A}\ }\textbf {\bibinfo {volume} {93}},\ \bibinfo {pages}
  {063819} (\bibinfo {year} {2016})}\BibitemShut {NoStop}%
\bibitem [{\citenamefont {Pelc}\ \emph {et~al.}(2012)\citenamefont {Pelc},
  \citenamefont {Yu}, \citenamefont {{De Greve}}, \citenamefont {McMahon},
  \citenamefont {Natarajan}, \citenamefont {Esfandyarpour}, \citenamefont
  {Maier}, \citenamefont {Schneider}, \citenamefont {Kamp}, \citenamefont
  {H{\"{o}}fling}, \citenamefont {Hadfield}, \citenamefont {Forchel},
  \citenamefont {Yamamoto},\ and\ \citenamefont {Fejer}}]{Pelc2012}%
  \BibitemOpen
  \bibfield  {author} {\bibinfo {author} {\bibfnamefont {J.~S.}\ \bibnamefont
  {Pelc}}, \bibinfo {author} {\bibfnamefont {L.}~\bibnamefont {Yu}}, \bibinfo
  {author} {\bibfnamefont {K.}~\bibnamefont {{De Greve}}}, \bibinfo {author}
  {\bibfnamefont {P.~L.}\ \bibnamefont {McMahon}}, \bibinfo {author}
  {\bibfnamefont {C.~M.}\ \bibnamefont {Natarajan}}, \bibinfo {author}
  {\bibfnamefont {V.}~\bibnamefont {Esfandyarpour}}, \bibinfo {author}
  {\bibfnamefont {S.}~\bibnamefont {Maier}}, \bibinfo {author} {\bibfnamefont
  {C.}~\bibnamefont {Schneider}}, \bibinfo {author} {\bibfnamefont
  {M.}~\bibnamefont {Kamp}}, \bibinfo {author} {\bibfnamefont {S.}~\bibnamefont
  {H{\"{o}}fling}}, \bibinfo {author} {\bibfnamefont {R.~H.}\ \bibnamefont
  {Hadfield}}, \bibinfo {author} {\bibfnamefont {A.}~\bibnamefont {Forchel}},
  \bibinfo {author} {\bibfnamefont {Y.}~\bibnamefont {Yamamoto}}, \ and\
  \bibinfo {author} {\bibfnamefont {M.~M.}\ \bibnamefont {Fejer}},\ }\href
  {\doibase 10.1364/OE.20.027510} {\bibfield  {journal} {\bibinfo  {journal}
  {Optics Express}\ }\textbf {\bibinfo {volume} {20}},\ \bibinfo {pages}
  {27510} (\bibinfo {year} {2012})}\BibitemShut {NoStop}%
\bibitem [{\citenamefont {Zaske}\ \emph {et~al.}(2011)\citenamefont {Zaske},
  \citenamefont {Lenhard},\ and\ \citenamefont {Becher}}]{Zaske2011}%
  \BibitemOpen
  \bibfield  {author} {\bibinfo {author} {\bibfnamefont {S.}~\bibnamefont
  {Zaske}}, \bibinfo {author} {\bibfnamefont {A.}~\bibnamefont {Lenhard}}, \
  and\ \bibinfo {author} {\bibfnamefont {C.}~\bibnamefont {Becher}},\ }\href
  {\doibase 10.1364/OE.19.012825} {\bibfield  {journal} {\bibinfo  {journal}
  {Optics Express}\ }\textbf {\bibinfo {volume} {19}},\ \bibinfo {pages}
  {12825} (\bibinfo {year} {2011})}\BibitemShut {NoStop}%
\bibitem [{\citenamefont {Zhao}\ \emph {et~al.}(2009)\citenamefont {Zhao},
  \citenamefont {Chen}, \citenamefont {Bao}, \citenamefont {Strassel},
  \citenamefont {Chuu}, \citenamefont {Jin}, \citenamefont {Schmiedmayer},
  \citenamefont {Yuan}, \citenamefont {Chen},\ and\ \citenamefont
  {Pan}}]{Zhao2009}%
  \BibitemOpen
  \bibfield  {author} {\bibinfo {author} {\bibfnamefont {B.}~\bibnamefont
  {Zhao}}, \bibinfo {author} {\bibfnamefont {Y.-A.}\ \bibnamefont {Chen}},
  \bibinfo {author} {\bibfnamefont {X.-H.}\ \bibnamefont {Bao}}, \bibinfo
  {author} {\bibfnamefont {T.}~\bibnamefont {Strassel}}, \bibinfo {author}
  {\bibfnamefont {C.-s.}\ \bibnamefont {Chuu}}, \bibinfo {author}
  {\bibfnamefont {X.-M.}\ \bibnamefont {Jin}}, \bibinfo {author} {\bibfnamefont
  {J.}~\bibnamefont {Schmiedmayer}}, \bibinfo {author} {\bibfnamefont {Z.-S.}\
  \bibnamefont {Yuan}}, \bibinfo {author} {\bibfnamefont {S.}~\bibnamefont
  {Chen}}, \ and\ \bibinfo {author} {\bibfnamefont {J.-W.}\ \bibnamefont
  {Pan}},\ }\href {\doibase 10.1038/nphys1153} {\bibfield  {journal} {\bibinfo
  {journal} {Nature Physics}\ }\textbf {\bibinfo {volume} {5}},\ \bibinfo
  {pages} {95} (\bibinfo {year} {2009})}\BibitemShut {NoStop}%
\end{thebibliography}

%

\onecolumngrid
\begin{appendix}
\vspace{40pt}
\begin{center}
\textbf{{\large Supplemental Material}}
\end{center}
\vspace{30pt}
\twocolumngrid

\subsection{TEORETICAL MODEL FOR THE SECOND-ORDER CROSS CORRELATION FUNCTION}\label{SWdephasing}
As mentioned in the main text, the second order cross-correlation function between the frequency converted write photons and the read photons, is related to the photon detection probabilities as $g^{(2)}_{cw,r}=p_{cw,r}/(p_{cw}p_r)$. During the frequency conversion process, the write photons experience two kinds of imperfections: the first one is imperfect transmission and the second one is that they are mixed with noise photons (coming mainly from residual pump light and detector dark counts). Considering these two effects, we can rewrite the photon detection probabilities as  $p_{cw}=\eta_\mathrm{QFC}p_{w}+p_N$ and $p_{cw,r}=\eta_\mathrm{QFC}p_{w,r}+p_Np_r$. In these expressions $\eta_\mathrm{QFC}$ is the total efficiency of the quantum frequency conversion device (QFCD), $p_w$ is the probability that a write photon arrives at the input of the QFCD and $p_N$ is the probability to detect a noise photon. These effects can be included in the expression of the cross-correlation function.

\ba
g^{(2)}_{cw,r}=\frac{\eta_{\mathrm{QFC}}p_{w,r}+p_Np_r}{(\eta_{\mathrm{QFC}}p_{w}+p_N)p_r}=\frac{\frac{p_{w,r}}{p_wp_r}+\frac{p_N}{\eta_{\mathrm{QFC}}p_w}}{1+\frac{p_N}{\eta_{\mathrm{QFC}}p_w}} \label{eq:1}
\ea

From the previous expression we can identify two terms. The first one is the cross-correlation of the fields without any frequency conversion $g^{(2)}_{w,r}=p_{w,r}/(p_wp_r)$ and the other one is the signal-to-noise-ration of the frequency converted photons $\mathrm{SNR} = (p_{cw}-p_N)/p_N = \eta_{\mathrm{QFC}}p_w/p_N$. Introducing this terms in eq.~(\ref{eq:1}) leads to the expression of eq.~(1) from the main text

\ba
g^{(2)}_{cw,r}=\frac{g^{(2)}_{w,r}+\mathrm{SNR}^{-1}}{1+\mathrm{SNR}^{-1}}
\ea

\subsection{MEASUREMENTS WITHOUT FREQUENCY CONVERSION}\label{SWdephasing}

\begin{figure}[htbp]
\includegraphics[width=7.5 cm]{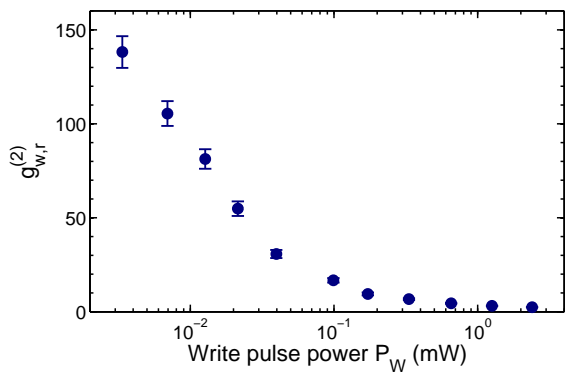}
\caption{Second order cross-correlation function of the write and read photon fields, without frequency conversion.}
\label{figS1}
\end{figure}

The blue shaded area in Fig. 3 of the main text shows the cross-correlation values that we expect after the frequency conversion of the write photons. As mentioned in the previous section, the way we obtain the expected $g^{(2)}_{cw,r}$ is by considering the cross-correlation values without the frequency conversion ($g^{(2)}_{w,r}$) and the imperfections that the QFC process would introduce.  The measured values of $g^{(2)}_{w,r}$ are shown in Fig.~\ref{figS1}. In order to take the data in the correct conditions (e.g. without additional noise), the write photons were filtered with a Fabry-Perot cavity with similar characterisitics as the one used for the read photons (see main text).

\subsection{THEORETICAL MODEL FOR THE DATA AS A FUNCTION OF THE STORAGE TIME}\label{SWdephasing}
Due to collective interference, spin-waves can be converted with a high efficiency into read photons emitted in a particular transition and direction. The ratio between the directional emission and the random emission (e.g. spontaneous emission) depends, among other factors, on the preservation of the spin-wave coherence. In our experiment this coherence is mainly affected by the random motion of the atoms, which induces a spin-wave dephasing that decreases the ratio between directional and random emission when the storage time increases. The effect of this dephasing on $\eta_{ret}$ and $g^{(2)}_{cw,r}$ can be seen in Fig.~4 of the main text.\\

 In order to find theoretical expressions for these quantities depending on the storage time, we start by writing them as a function of the photon detection probabilities $\eta_{ret}=p_{cw,r}/p_w$ and $g^{(2)}_{cw,r}=p_{cw,r}/(p_{cw}p_{r})$. Since the ratio between the read photon directional and random emission will depend on the storage time, we can write $p_{r}(t)=p_{r}^{dir}(t)+p_{r}^{rand}(t)$ and $p_{cw,r}(t)=p_{cw,r}^{dir}(t)+p_{cw}p_{r}^{rand}(t)$. The ratio of directional emission can be characterized by the intrinsic retrieval efficiency $\eta^I_{ret}(t)$, while the random emission is proportional the total number of atoms in the $\ket{s}$ ground state ($N_s$), the branching ratio of the read photon atomic transition ($\xi_g=1/6$) and the read photon spatial mode solid angle ($\Delta\Omega_{r}$). Hence, we can write all the photon detection probabilities as:

\ba 
p_{cw}&=&p\eta_{cw} \label{eq:3}\\
p_r(t) &=&p\eta^I_{ret}(t)\eta_{r}+N_s[1-\eta^I_{ret}(t)]\frac{\Delta\Omega_r}{4\pi}\xi_g\eta_r \label{eq:4}\\
p_{cw,r}(t) &=&p\eta^I_{ret}(t)\eta_{cw}\eta_{r}+p\eta_{cw}N_s[1-\eta^I_{ret}(t)]\frac{\Delta\Omega_r}{4\pi}\xi_g\eta_r \label{eq:5}
\ea\

where $p$ is the probability to create a spin-wave together with a write photon in the coupled spatial mode and $\eta_{cw(r)}$ are the write (read) photon total detection efficiencies. In the previous equations we have taken into account non-unity detection efficiencies, but we have not considered any noise. The reason is that we experimentally measured that the probability to detect a noise count for the data in Fig. 4 is much lower than the probability to detect a real photon. The noise detection probabilities for the write and read photon modes are $p_{Nw}=2.30\cdot10^{-5}$ and $p_{Nr}=7.8 \cdot10^{-5}$, respectively.\\

Since write photons and excitations in state $\ket{s}$ are created in pairs, $N_s$ can be obtained from $p$ and the write photon spatial mode solid angle $\Delta\Omega_{w}$, with the expression $N_s=p4\pi/\Delta\Omega_{w}$. In our experiment the solid angle of the write and read photonic modes are the same ($\Delta\Omega_{w}=\Delta\Omega_{r}$).\\

The storage time dependence of the directional emission will be represented by $\eta^I_{ret}(t)$ wich is proportional to the overlap between the state of the dephased spin-wave with the unperturbed one: $\eta^I_{ret}(t)\propto\left|\left<\Psi(0)|\Psi(t)\right>\right|^2$. As explained in literature \cite{Zhao2009,Albrecht2015}, in the presence of random atomic motion dephasing, the spin-wave state evolves with time as

\ba
\ket{\Psi(t)}=\frac{1}{N}\sum_je^{i(\bold{k_W}-\bold{k_w})(\bold{x_j}+\bold{v_j}t)}\ket{g_1...s_j...g_N}
\ea 
 where $\bold{k_{W(w)}}$ is the wave vector of the write pulse (write photon), $N$ is the total number of atoms, $\bold{x_j}$ is the position of atom j at the moment of the write process and $\bold{v_j}$ its velocity.
Assuming a Maxwell-Boltzmann distribution for the atomic velocities, the temporal dependence of the intrinsic retrieval efficiency becomes $\eta^I_{ret}(t)=\eta^I_{ret}e^{-t^2/\tau^2}$ with $\tau=\sqrt{m/(k_BT\Delta k^2)}$, where $m$ is the atomic mass, $k_B$ is the Boltzmann constant, $T$ is the atomic temperature and $\Delta k=|\bold{k_W}-\bold{k_w}|$ \cite{Albrecht2015}.\\

Introducing the previous expressions in equations (\ref{eq:3})-(\ref{eq:5}), we can finally write the theoretical formulas used for the fits in Fig.4 of the main text. 

\ba
\eta_{ret}=\eta_r\left[\eta^I_{ret}e^{-t^2/\tau^2}(1-p\xi_g)+p\xi_g\right]\\
g^{(2)}_{cw,r}=1+\frac{\eta^I_{ret}e^{-t^2/\tau^2}(1-p)}{p\left[\eta^I_{ret}e^{-t^2/\tau^2}(1-\xi_g)+\xi_g\right]}
\ea 

\end{appendix}

\end{document}